\documentclass[10pt,a4paper,twoside]{article}
\usepackage{epsfig}
\usepackage{baltlat6}
\usepackage{array}
\usepackage{here}
\begin{document}
\ \
\vspace{0.5mm}
\setcounter{page}{253}
\vspace{8mm}

\titlehead{Baltic Astronomy, vol.\,17, 253--275, 2008}

\titleb{2MASS TWO-COLOR INTERSTELLAR REDDENING LINES IN\\ THE INNER
GALAXY}

\begin{authorl}
\authorb{V. Strai\v{z}ys}{} and
\authorb{V. Laugalys}{}
\end{authorl}

\moveright-3.2mm
\vbox{
\begin{addressl}
\addressb{}{Institute of Theoretical Physics and Astronomy, Vilnius
University,\\  Go\v{s}tauto 12, Vilnius LT-01108, Lithuania;
straizys@itpa.lt}
\end{addressl}   }

\submitb{Received: 2008 November 2; accepted: 2008 December 15}

\begin{summary} The slopes of interstellar reddening lines in the 2MASS
$J$--$H$ vs.  \hbox{$H$--$K_s$} diagram for 26 areas in the inner Galaxy
(from Vulpecula to Centaurus) are determined.  For this aim we use the
red-clump giants located inside and behind spiral arms, or behind dense
dust clouds of the Local arm.  In most of the investigated directions
the ratio $E_{J-H}/E_{H-K_s}$ is found to be between 1.9 and 2.0, taking
the stars with the extinction $A_V$\,$<$\,12 mag.  The stars with larger
extinction deviate down from the reddening lines corresponding to less
reddened stars.  Probably, this is related to the curvature of reddening
lines due to the band-width effect.  However, some of the deviating
stars may be heavily reddened oxygen- and carbon-rich AGB stars (giants
of the latest M subclasses or N-type carbon stars), and
pre-main-sequence objects (YSOs).  \end{summary}


\resthead{2MASS two-color interstellar reddening lines in the inner
Galaxy}
{V. Strai\v{z}ys, V. Laugalys}

\sectionb{1}{INTRODUCTION}

In the previous paper (Strai\v{z}ys et al. 2008, hereafter Paper I) we
have shown that in the direction of the North America and Pelican
Nebulae and the Cyg OB2 association the ratio of color excesses
$E_{J-H}/E_{H-K_s}$ is close to 2.0, i.e., it is larger than the
`typical' value of 1.7--1.8 which follows from the standard interstellar
extinction law.  Similar values have been obtained by Racca et al.
(2002) in the Coalsack and Hoffmeister et al.  (2008) in the M\,17
nebula.  This result stimulated verification of the ratio of 2MASS color
excesses in other directions, including dust clouds in the Local arm and
more distant clouds.  In the present paper we report the results
obtained in the direction of the inner Galaxy, between $\ell$ = 60\degr\
(Vulpecula) and 310\degr\ (Centaurus).  The covered areas include both
distant clouds in the Sagittarius, Scutum and Norma arms and the
Galactic bulge/bar, and the Gould Belt clouds close to the Sun in
Aquila, Serpens, Ophiuchus, Scorpius, Sagittarius, Corona Australis and
Lupus.

\newpage

\sectionb{2}{THE INVESTIGATED AREAS}

The areas for the determination of the ratio of color excesses were
selected in Region 1 of the Dobashi et al.  (2005) atlas.  Two types
of areas were taken:  (1) the areas located on the Galactic equator
every 10 degrees; (2) the densest parts of dust clouds of the Local arm
belonging to the Gould Belt.  All areas are round with a diameter of
1\degr.  The lists of the areas are given in Tables 1 and 2. In Figure 1
the areas are plotted on the map of Region 1 (Dobashi et al. 2005).

The infrared sources were selected from the 2MASS point-source
catalog (Cutri et al. 2006; Skrutskie et al. 2006) taking the $J$, $H$
and $K_s$ magnitudes with the errors $\leq 0.03$ mag.

%%%%%%%%%%%%%%%%%%%%%%%%%%%%%  FIGURE 1

\begin{figure}[H]
\vbox{
\centerline{\psfig{figure=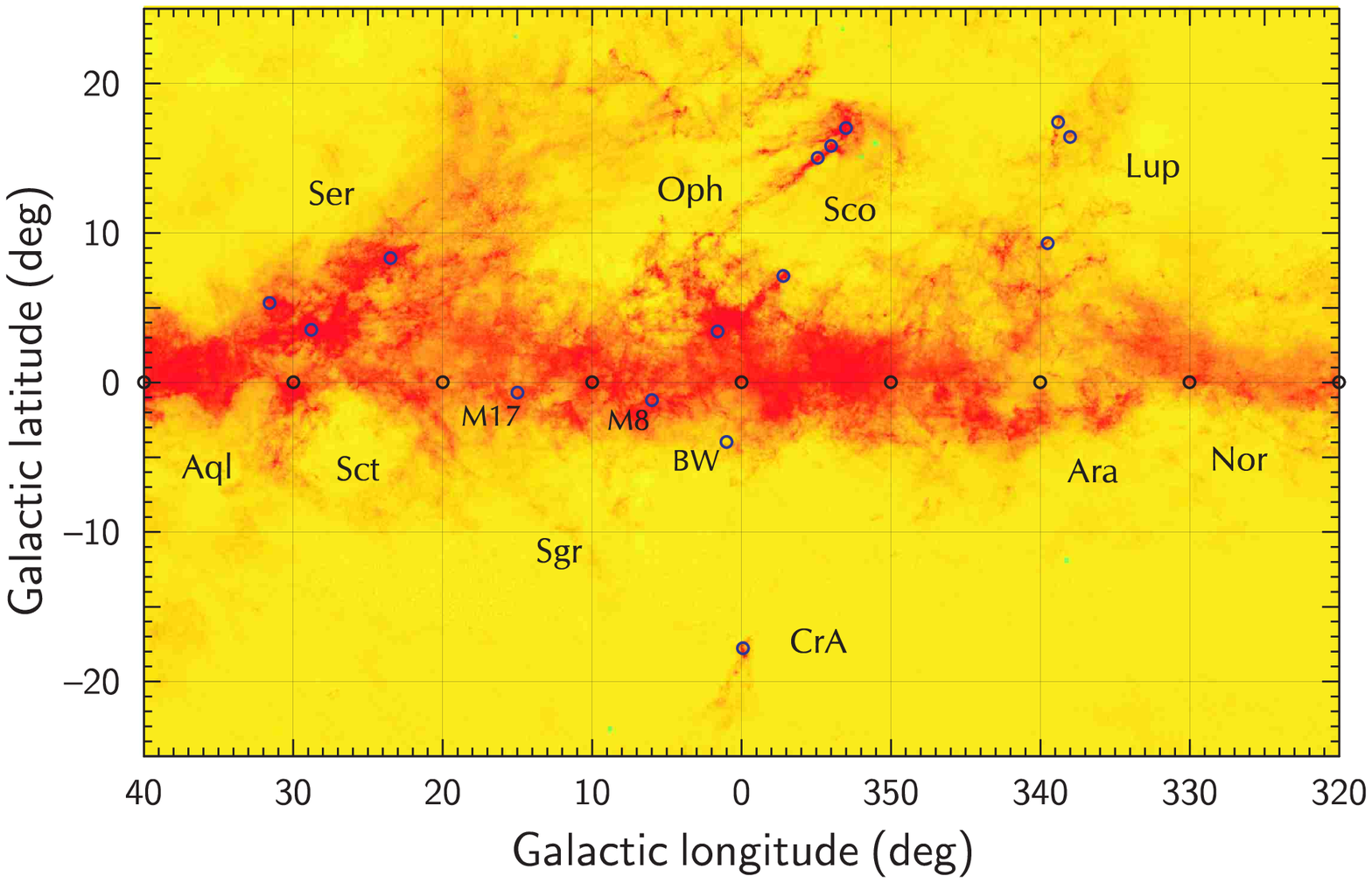,width=125mm,angle=0,clip=}}
\vspace{1mm}
\captionb{1}
{The investigated areas plotted on the map of dust clouds from Dobashi
et al. (2005). Two areas on the Galactic equator at $\ell$ = 60\degr\ and
310\degr\ are outside the map. BW means Baade's Window.}
}
\end{figure}

%%%%%%%%%%%%%%%%%%%%%%%%%%%%%%%%%%%%%%%%%%%%%%%%
\vskip-2mm

\sectionb{3}{THE METHOD}

The majority of stars, located behind a dense dust cloud or a group of
clouds, in the $J$--$H$ vs.~$H$--$K_s$ diagram form a feature similar
to a cometary tail.  The main constituent of the tail are K and M giants
located at different distances and reddened by dust clouds distributed
along the line of sight.  Among these stars, the red clump giants
(hereafter RCGs) of spectral type $\sim$\,K2\,III are most abundant.
Their absolute magnitude in the $K_s$ passband, $M_{K_s}$, is close to
--1.6 (Alves 2000; Grocholski \& Sarajedini 2002; Groenewegen 2008).

%\footnote{~The red clump giants are Population I analogs of the red
%horizontal branch of globular clusters.  They burn helium at the core
%and hydrogen in a spherical shell.}

The upper edge of this tail usually is much sharper than the lower edge
where a large variety of objects with different interstellar reddenings
can be present.  These objects can belong to the following types:  (1)
AGB stars, including the coolest M-type giants, oxygen- and carbon-rich
long-period variables, OH/IR stars; \hbox{(2) young} stellar objects
(YSOs), including Orion-type variables and Herbig Ae/Be stars in
different stages of evolution; (3) ordinary Be-type stars with gas
envelopes; (4) point-like galaxies and quasars.  Sometimes these objects
form the second tail which usually is broader than the upper one (see
Strai\v zys \& Laugalys 2007).  If the dust cloud (or other clouds
behind it) does not show star-forming activity, the YSOs are absent.
Galaxies, quasars and Be stars usually are not numerous.  Consequently,
the objects scattered along the lower edge of the main `tail' in many
cases are cool AGB stars only.

For determining the reddening line slope (i.e., the ratio
$E_{J-H}/E_{H-K_s}$) we applied the method described in Paper I, Section
5. It is based on using the tail formed by reddened RCGs, scattered
around a line crossing their intrinsic position at $J$--$H$ = 0.60 and
$H$--$K_s$ = 0.15.  The only difference is that in Paper I we used a
little different intrinsic position of RCGs determined for the clump
stars in the open cluster M\,67.  Now it is evident that M\,67 RCGs are
too hot (spectral types G8--K0 III) comparing to the majority of the
field RCGs which are mostly of spectral types closer to K2\,III.

Although the number of RCGs exceeds considerably the number of stars
which in the color-magnitude diagram lie on the sequence of red giants,
most of them in the $J$--$H$ vs.~$H$--$K_s$ diagram form a common
reddening line since their intrinsic line almost coincides with the
direction of the reddening line.  Only the giants of spectral types
later than M5\,III (most of them are asymptotic-branch giants) deviate
down from the G8--M5 giant sequence (see Figure 4b).  Consequently, we
are free to consider that all giants from G8 to M5 in the $J$--$H$
vs.~$H$--$K_s$ diagram form a common reddening line of red giant branch.

To make the estimation of the reddening line slope easier, in the
$J$--$H$ vs.~$H$--$K_s$ diagram of each area a fan of reddening lines
of RCGs with the slopes between 1.7 and 2.2 were drawn.  The optimum
slope for each area was determined by taking the average of slopes of
individual reddening lines for stars with color indices \hbox{$J$--$H$}
between 1.8 and 2.2.  We have avoided cooler stars since the reddening
lines exhibit curvature due to the band-width effect.

In the following we accept the location of spiral arms according to the
Vallee (2005) four-arm model.  The recent model of Benjamin et al.
(2008) accepting two star-rich arms and two star-poor arms, based on the
{\it Spitzer} GLIMPSE project, does not change the positions of spiral
arms close to the Sun.  Consequently, the main results of our study
remain valid in both models.  We accept that in the Galactic center
direction the heliocentric distance of the Sagittarius arm is 1--2 kpc
and of the Scutum arm it is 2.5--3.5 kpc.

\sectionb{4}{RESULTS}

A general feature in $J$--$H$ vs.~$H$--$K_s$ diagrams for most of the
areas is an uneven distribution of stars along the reddening line of
red giants.  The presence of clumps or density jumps at certain color
excesses is related to the distribution of interstellar dust clouds
along the line of sight.  Usually, the dust density is the largest
inside spiral arms.  The distribution of density jumps in the $J$--$H$
vs.~$H$--$K_s$ diagram can be better understood analyzing the
distribution of stars in corresponding $K_s$ vs.  $H$--$K_s$
color-magnitude diagrams (hereafter CMD).  In Figures 2 and 3 two
typical diagrams for the areas on the Galactic equator at $\ell$ =
50\degr\ (in Aquila) and $\ell$ = 2.5\degr\ (close to the Galactic
center) are presented.

%%%%%%%%%%%%%%%%%%%%%%%%%  FIGURES 2 AND 3

\begin{figure}[H]
\centerline{\hbox{
\parbox{61mm}{\psfig{figure=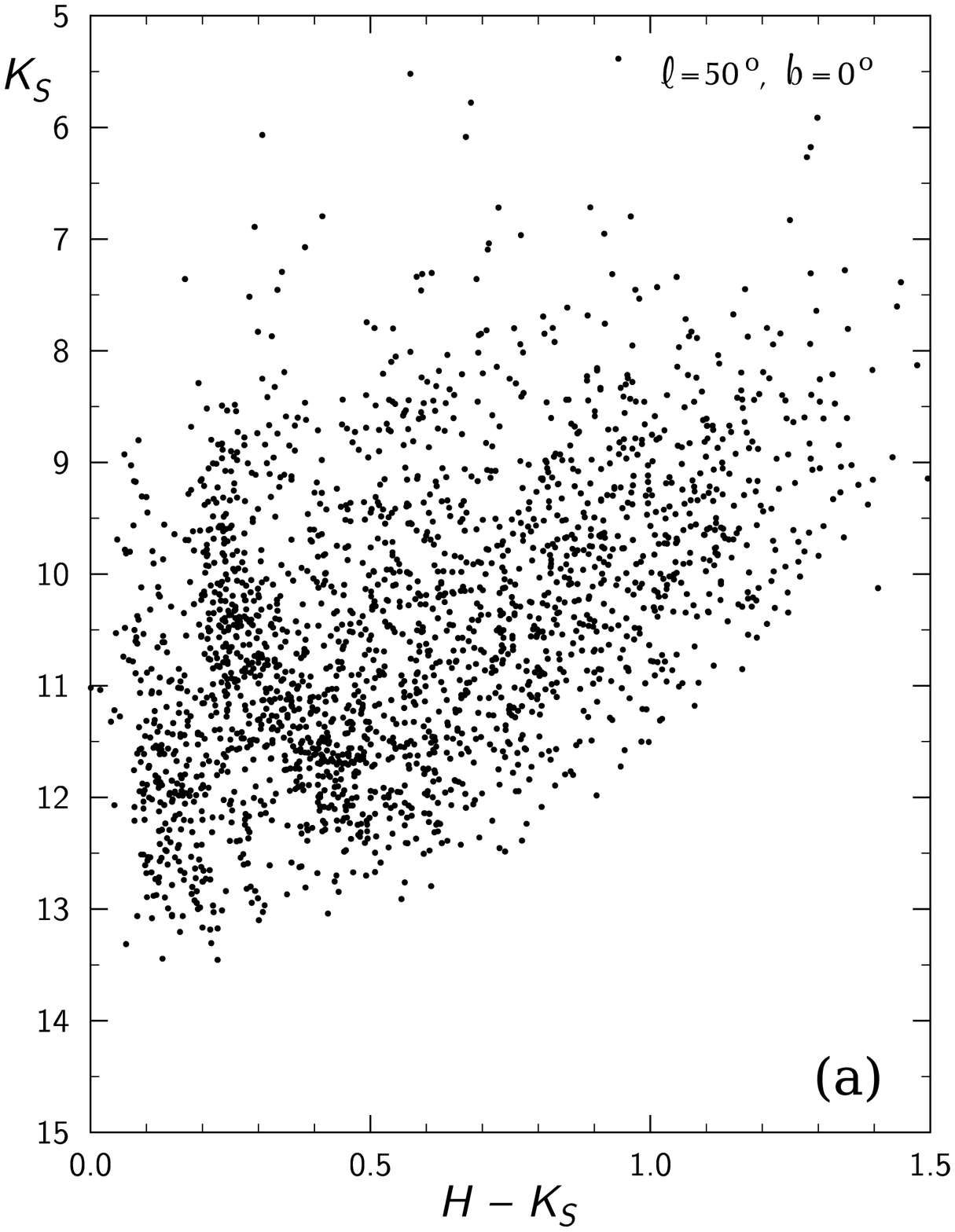,angle=0,width=61truemm,clip=}}
\hskip3mm
\parbox{61mm}{\psfig{figure=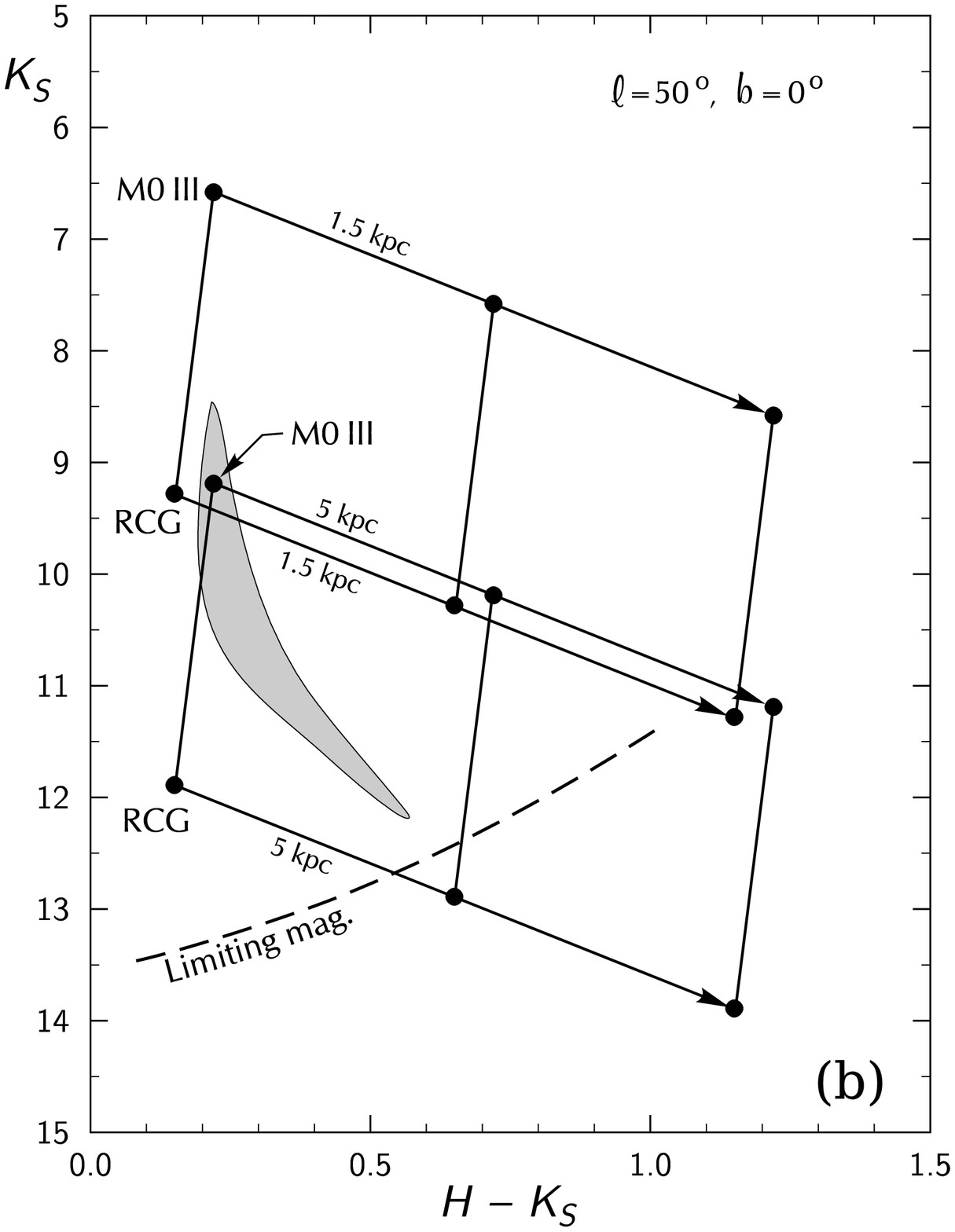,angle=0,width=61truemm,clip=}}}}
\vskip2mm
\captionb{2}{Color-magnitude diagram $K_s$ vs.~$H$--$K_s$ and its
interpretation in the direction of $\ell$ = 50\degr, $b$ = 0\degr\ in
Aquila.}
\vskip7mm
\centerline{\hbox{
\parbox{61mm}{\psfig{figure=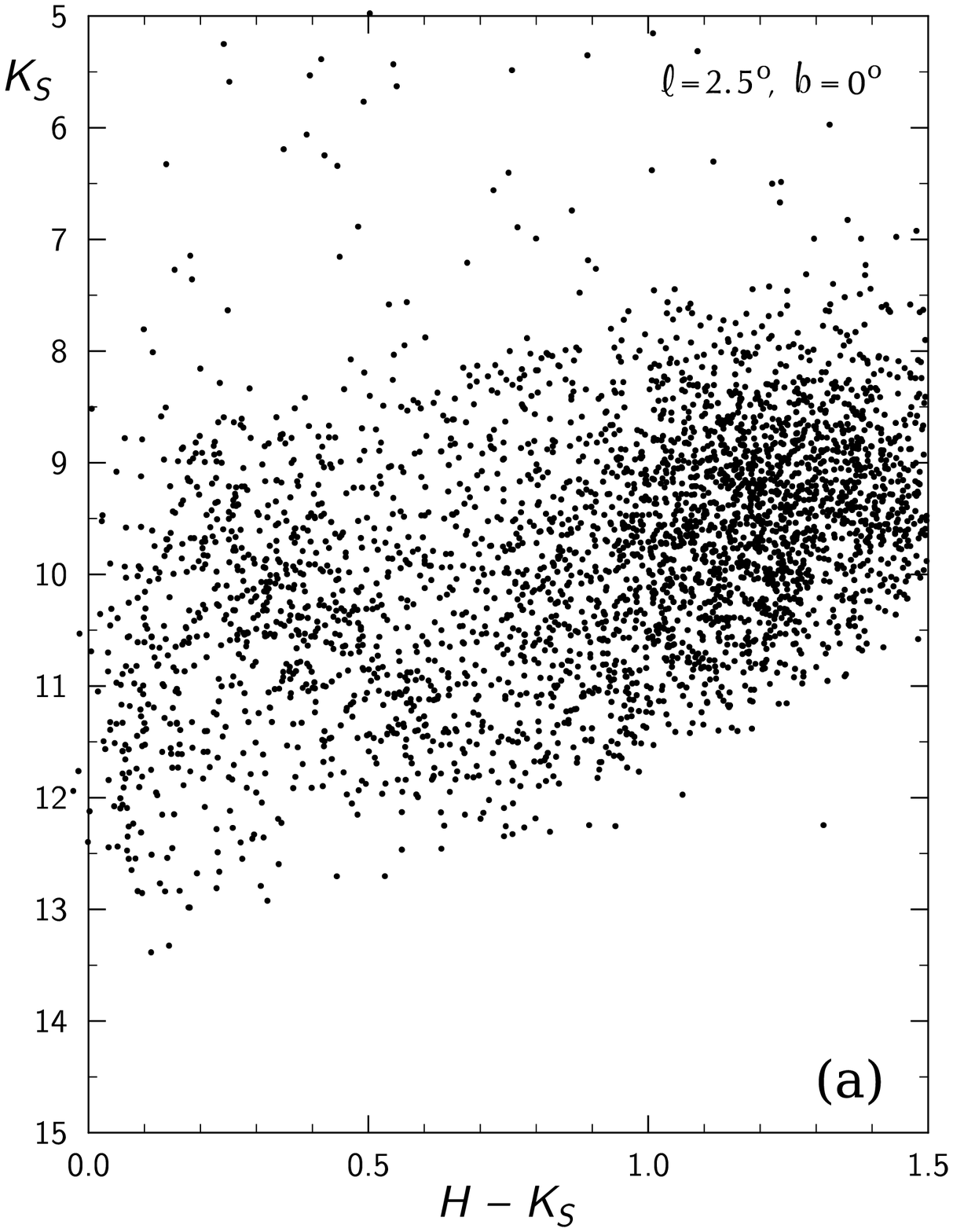,angle=0,width=61truemm,clip=}}
\hskip3mm
\parbox{61mm}{\psfig{figure=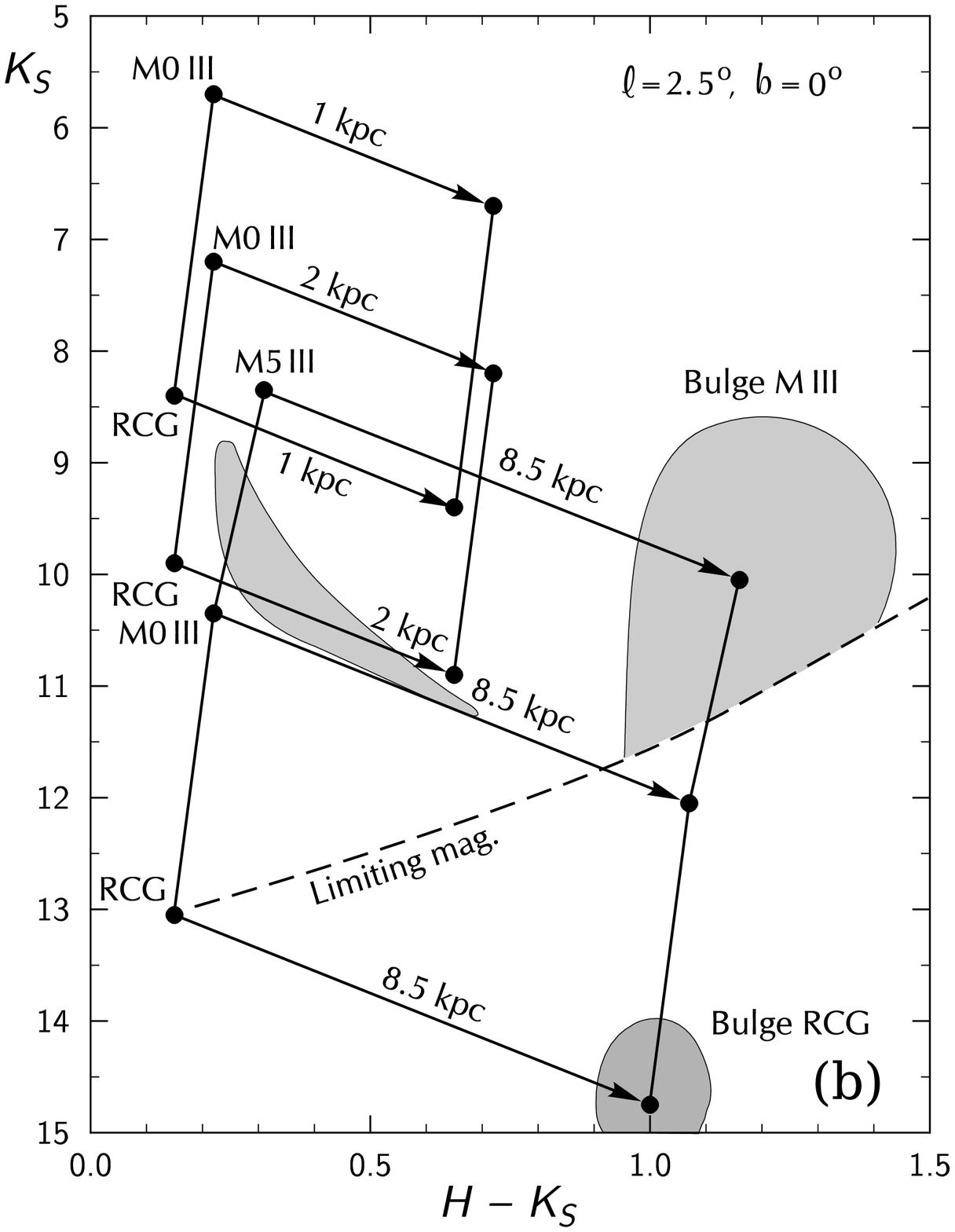,angle=0,width=61truemm,clip=}}}}
\vskip2mm
\captionb{3}{Color-magnitude diagram $K_s$ vs.~$H$--$K_s$ and its
interpretation in the direction of $\ell$ = 2.5\degr, $b$ = 0\degr\ at
the Galactic center.}

\end{figure}
\newpage

\subsectionb{4.1}{Color-magnitude diagram along the Sagittarius arm}

In the area at $\ell$ = 50\degr\ within the heliocentric distance of
2--5 kpc the line of sight runs along the Sagittarius arm.  In Figure 2
the left panel is plotted for the 2MASS data while the right panel gives
a schematic explanation of the features observed.  The positions of the
unreddened and reddened by $E_{H-K_s}$ = 0.5 and 1.0 stars for RCGs and
M0\,III giants at distances of 1.5 and 5 kpc, joined by solid lines, are
shown.  These color excesses correspond to $A_V$ = 8.5 and 17.0 mag.
The slope of the reddening line, or the ratio $A_{K_s} / E_{H-K_s}$ is
accepted to be 2.0.  The broken line at the bottom shows the limiting
2MASS $K_s$ magnitudes at which the accuracy of photometry in all three
passbands is $\leq$\,0.03 mag.

The most conspicuous feature in the diagram is the sequence of stars
running from ($K_s$, $H$--$K_s$) = (8.5, 0.22) to (12.2, 0.6).  This
sequence corresponds to RCGs of spectral types K0--K3 III located in
front of the Sagittarius arm (upper end) and within the arm.  In this
direction the space density of RCGs with distance should be more or less
constant, but due to increasing volume of space at larger distances the
surface density of stars increases.  The width of the RCG sequence is
defined by the presence of cloud clumps and windows in the arm.

In Figure 2 the RCG sequence appears at a distance of about 1 kpc where
the cone with a diameter of 1\degr\ counts up a sufficient number of
RCGs at ($K_s$, $H$--$K_s$) = (8.5, 0.22).  These stars are already
reddened with $E_{H-K_s} \approx 0.07$ ($A_V$ = 1.2~mag) by the Aquila
Rift dust clouds located at a distance of $\sim$\,200--300 pc (Strai\v
zys et al. 1996, 2003; Eiroa et al. 2008; Prato et al. 2008).  Between
the Local and the Sagittarius arms the reddening should be small.
Therefore at the beginning the RCG sequence runs almost vertically down
due to increasing distance.  At $\sim$\,2 kpc ($K_s$ = 10.2) the
sequence turns to  right forming a curve running down to the end at
($K_s$, $H$--$K_s$) = (12.2, 0.6) where the line of sight leaves the
Sagittarius arm.  At larger distances the line should fall more or less
vertically down before the Perseus arm is reached.  However, this
extension of the line is not seen due to the limiting magnitude.

The dust in a spiral arm shifts along the reddening line not only RCGs,
but all the red giant sequence, which extends both down and up from the
clump (except RCGs, in Figure 2b we show the positions of M0\,III
stars).  The K3--M5 III stars of the Sagittarius arm fill the region of
the CMD above the RCG sequence, with \hbox{$H$--$K_s$} approximately
between 0.25 and 0.7.  M-type giants of late subclasses, located between
the Sagittarius and Perseus arms, due to their high luminosity in the
$K_s$ passband, also should be present in Figure 2, right of
$H$--$K_s$\,$\approx$\,0.7.

The stars at the left lower corner of the CMD with $H$--$K_s$\,$<$\,0.2
are unreddened or slightly reddened main sequence stars of spectral
types A-F-G located up to a distance of $\sim$\,500 pc from the Sun.

\subsectionb{4.2}{Color-magnitude diagram in the direction of the
Galactic bulge}

Figure 3a shows the $K_s$ vs.~$H$--$K_s$ diagram at ($\ell$, $b$) =
(2.5\degr, 0\degr), i.e., for a region close to the Galactic center.  We
have avoided the area placed directly on the center since in it, due to
very large interstellar extinction, the distribution of stars is not
typical for other areas in the direction of the bulge.  In the bulge
direction the line of sight crosses the Sagittarius arm and other arms
almost perpendicularly.

Figure 3b gives the interpretation of the distribution of stars.  Again,
as in Figure 2, we see the sequence of RCGs which starts more or less at
the same $K_s$ magnitude and $H$--$K_s$ color with small reddening,
$E_{H-K_s}$ = 0.25, caused by dust clouds in the Gould Belt (probably,
an extension of the Pipe Nebula cloud).  At a distance of 1 kpc the line
of sight enters the Sagittarius arm.  The RCG sequence is not so steep
as at 50\degr\ and runs to somewhat larger reddenings (up to $H$-$K_s$ =
0.7).  Both these facts mean that the dust concentration in the
Sagittarius arm in the direction of the Galactic center is larger than
along the arm at the 50\degr\ longitude.  Also, the RCG sequence in
Figure 3 is broader, and this may mean that the extinction in this
direction is more spotty.  Near the end of the RCG sequence, whose form
is defined by the dust distribution in the Sagittarius arm, the giants
in the Scutum arm should appear (at about 2.5 kpc).  However, in Figure
3 we do not see any signs of the Scutum arm, maybe due to proximity of
its RCGs to the limiting magnitude.

In comparison with Figure 2, a new feature appears in the right side of
Figure 3 -- between $K_s$ = 8--11 and $H$--$K_s$ = 1.0--1.5 there is a
large clump of stars belonging to the Galactic bulge.  As panel 3b
explains, this concentration is only a `tip of the iceberg', the long
red giant sequence of the bulge, with the reddening $E_{H-K_s}$ = 0.85,
or the extinction $A_V$ = 14.5 mag.  Dust clouds responsible for this
large reddening are located in the Local, Sagittarius, Scutum, Norma and
3-kpc arms.  RCGs on this sequence are located at $K_s$ =
14.75, i.e., far below the limiting magnitude.  The presence of the
clump at approximately this position was confirmed by Nishiyama et al.
(2006).  This is in accordance with the intrinsic RCG position at $K_s$
= 13 mag shown in Figure 3b.  This position of RCGs of the bulge was
confirmed by Tiede et al.  (1995), using {\it JHK} photometry in Baade's
Window.

Above the CMD limiting magnitude only M giants of the bulge are seen.
As follows from Figure 3b, many of them are situated higher than the
giants of spectral class M5\,III located at a distance of 8.5 kpc.  This
can be explained by a large radial extent of the bulge which starts at a
heliocentric distance of $\sim$\,5 kpc.  Also, a part of the bulge stars
are cooler and more luminous in $K_s$ than M5, most of them are AGB
long-period variables (Schultheis \& Glass 2001).  Their scatter in CMD
can be the result of their variability, presence of warm circumstellar
gaseous and dusty envelopes and different interstellar reddenings.

Using the DENIS and 2MASS data, the described M0--M6 giants of the bulge
with intrinsic $K_s$ magnitudes between 8 and 12 have been used by
Schultheis et al.  (1999) and Dutra et al.  (2002, 2003) for the
extinction investigation in the bulge.  The same method, applying the
$K$ vs.~$J$--$K$ diagram, was used earlier by Tiede et al.  (1995)
and Frogel et al.  (1999) in the NICMOS 3 (Las Campanas) system.

Figures 2 and 3 show that the presence of interstellar extinction and
reddening makes possible to identify giants of different spectral
classes belonging to the specific spiral arms and the bulge.  Without
reddening or with small reddening such possibility vanishes.  This is
illustrated by the CMD for Baade's Window, a relatively transparent
region ($A_V$\,$\approx$\,1.6 mag for the globular cluster NGC 6522 seen
in the center) in the direction of the Galactic bulge at $\ell$ =
1\degr, $b$ = --4\degr\ (Figure 4a).  In this diagram the giants located
at various distances and belonging to the disk and the bulge form a
single band.

%%%%%%%%%%%%%%%  Figure 4a and 4b, Baade's Window

\vbox{
\centerline{\psfig{figure=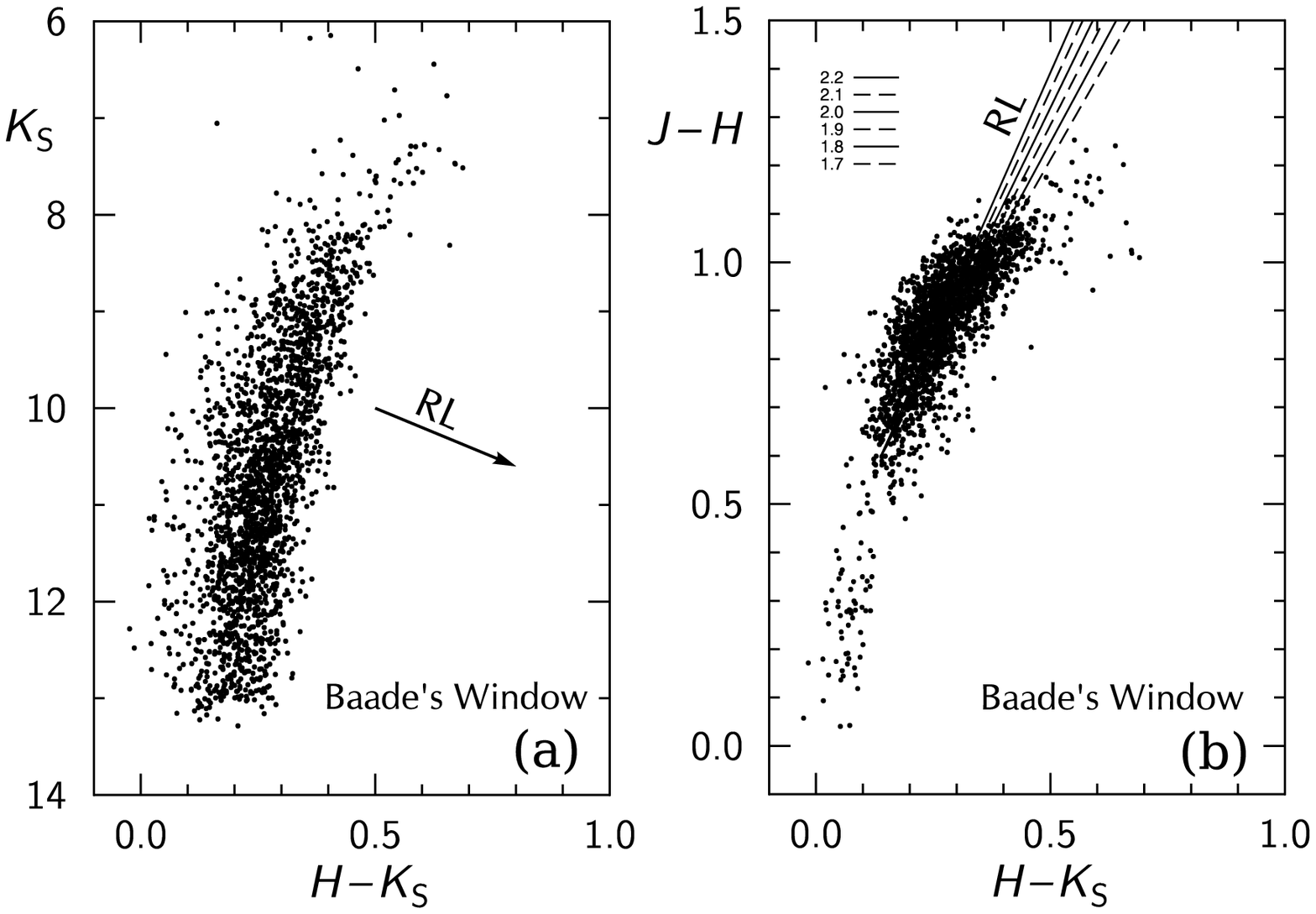,angle=0,width=125truemm,clip=}}
\captionb{4}{Color magnitude and two-color diagrams for Baade's Window
($\ell$ = 1\degr,\\ $b$ = --4\degr, diameter 15\arcmin).}
}
\vskip.5mm

%%%%%%%%%%%%%%%%%%%%%%%%%%%%%%%%%%  TABLE 1
\begin{table}[!h]
\begin{center}
\vbox{\small\tabcolsep=6pt
\parbox[c]{124mm}{\baselineskip=10pt
{\normbf\ \ Table 1.}{\norm\
The investigated areas located on the Galactic equator.  The
given mean ratios of color excesses correspond to red giant branch stars
at $J$--$H$\,$\approx$\,2.0.  The last column shows the number of stars
used to calculate the ratio of color excesses.
}}
\begin{tabular}{lrccr}
\tablerule
Area & \multicolumn{2}{c}{Center} & $E_{J-H}/E_{H-K_s}$     &  Number \\
  &   $\ell$\degr\ & $b$\degr\  & at $J$--$H$ = 2.0      &  of stars \\
\tablerule
Vul (60,0)     &  60.0  &  0.0 & $2.035\pm 0.158$ &  216~~~ \\
Aql (50,0)     &  50.0  &  0.0 & $1.992\pm 0.142$ &  313~~~ \\
Aql (40,0)     &  40.0  &  0.0 & $1.976\pm 0.157$ &  933~~~ \\
Aql/Sct (30,0) &  30.0  &  0.0 & $1.984\pm 0.156$ &  484~~~ \\
Sct (20,0)     &  20.0  &  0.0 & $1.977\pm 0.152$ &  709~~~ \\
Sgr (10,0)     &  10.0  &  0.0 & $1.935\pm 0.140$ &  557~~~ \\
Sgr (0,0)      &   0.0  &  0.0 & $2.018\pm 0.175$ &   72~~~ \\
Sco (350,0)    & 350.0  &  0.0 & $1.958\pm 0.147$ &  594~~~ \\
Sco/Ara (340,0)& 340.0  &  0.0 & $1.984\pm 0.168$ &  514~~~ \\
Nor (330,0)    & 330.0  &  0.0 & $1.949\pm 0.132$ &  566~~~ \\
Cir (320,0)    & 320.0  &  0.0 & $1.989\pm 0.145$ &  634~~~ \\
Cen (310,0)    & 310.0  &  0.0 & $2.000\pm 0.144$ &  354~~~ \\
\tablerule
\end{tabular}
}
\end{center}
\vskip-4mm
\end{table}

\subsectionb{4.3}{Two-color diagrams at the Galactic plane}

Two-color $J$--$H$ vs.~$H$--$K_s$ diagrams for the Galactic equator
areas between \hbox{$\ell$ =} 60\degr\ and 310\degr\ at each longitude
multiple to 10\degr\ are shown in Figures 5 (a--$\ell$).  The areas are
listed in Table 1 with the calculated mean slopes of their reddening

%%%%%%%%%%%%%%%%%  FIGURES 5

%\begin{figure}[!th]
\vbox{
\centerline{\psfig{figure=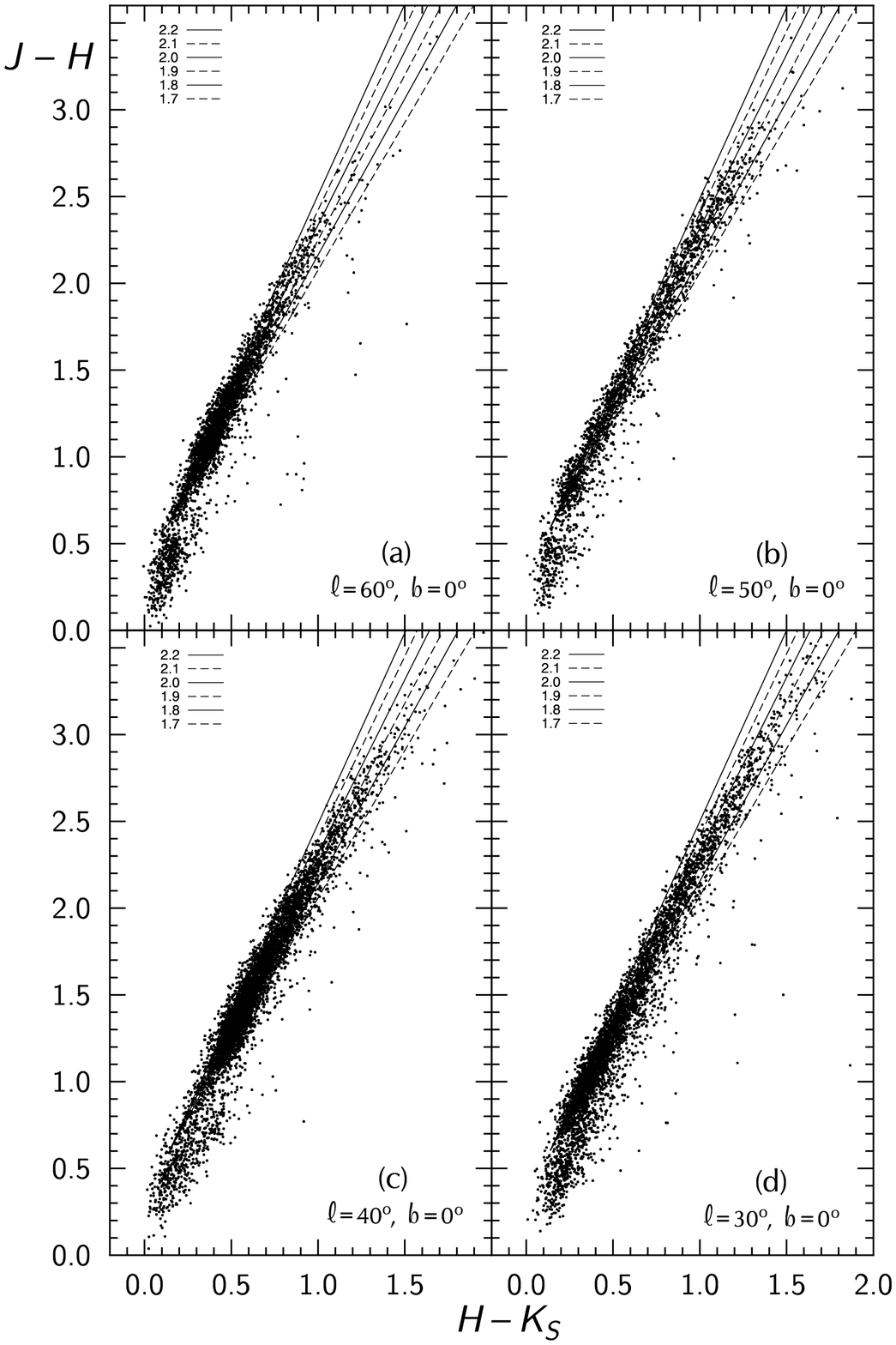,angle=0,width=120truemm,clip=}}
\vskip2mm
\captionb{5\,(a--d)}{Two-color diagrams for the areas on the Galactic equator
at $\ell$ = 60\degr, 50\degr, 40\degr and 30\degr. The six straight
lines are the reddening lines of RCGs with different slopes, explained
in the inserts.}
}
%\end{figure}

\vbox{
\centerline{\psfig{figure=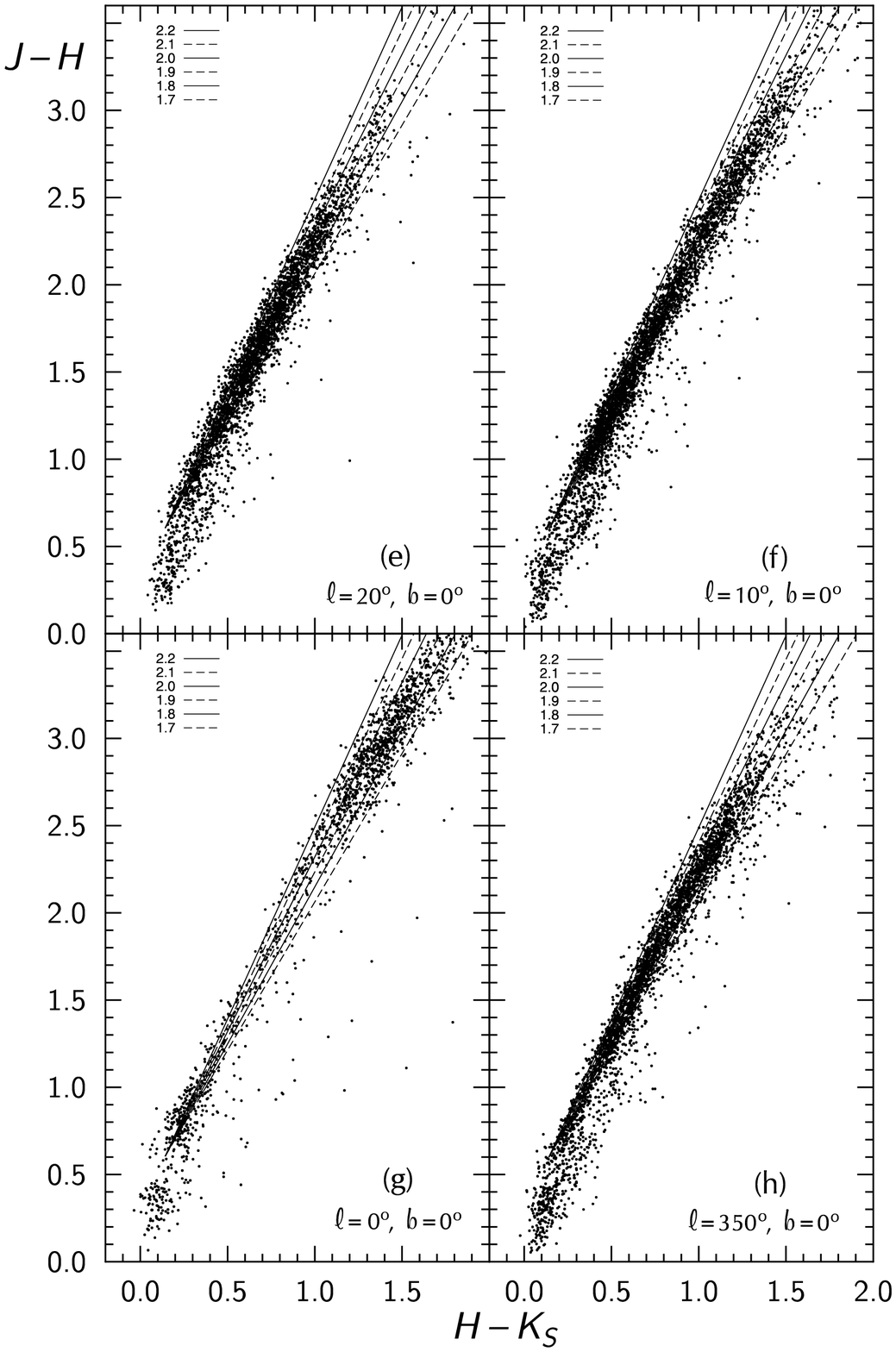,angle=0,width=120truemm,clip=}}
\vskip2mm
\captionb{5\,(e--h)}{The same as in Fig. 5 (a--d) but for the areas at $\ell$
= 20\degr, 10\degr, 0\degr\ (Galactic center) and 350\degr.}
}

\vbox{
\centerline{\psfig{figure=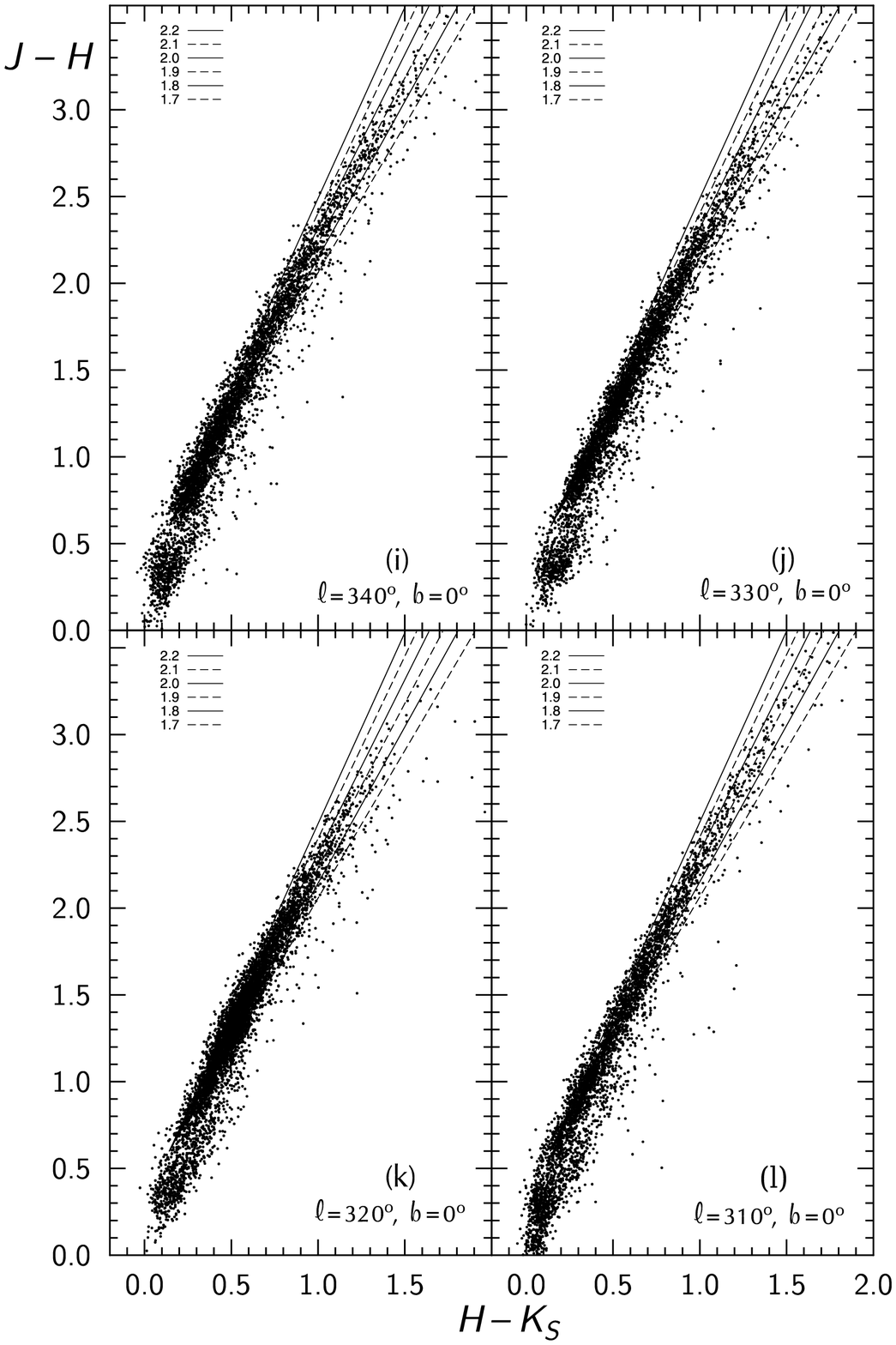,angle=0,width=120truemm,clip=}}
\vskip2mm
\captionb{5\,(i--l)}{The same as in Fig. 5 (a--d) but for the areas at $\ell$
= 340\degr, 330\degr, 320\degr\ and 310\degr.}
}

\noindent lines for
red giants with $J$--$H$ between 1.8--2.2.  When calculating slopes of
the reddening lines, the possible AGB stars and YSOs located below the
reddening line of the slope 1.6, have been excluded.  In the next we
will discuss these diagrams considering both the distribution of stars
along reddening lines and the ratios of color excesses in the given
directions.

\subsubsectionb{4.3.1}{$\ell$ = 60\degr, $b$ = 0\degr}

In this direction the line of sight runs along the inner edge of the
Local arm, where space is relatively transparent.  However, at
$\sim$\,400 pc it crosses a dust cloud (hereafter the Vulpecula cloud,
Neckel \& Klare 1980; Forbes 1985) and at 2--2.5 kpc a cloud inside the
Vul OB1 association (Dame \& Thaddeus 1985).  In the CMD of the area the
RCG sequence, corresponding to the Local arm, runs steeply down between
($K_s$, $H$--$K_s$) = (9.0, 0.2) and (11.0, 0.35).  At a distance of
5--6 kpc the line of sight enters the branching of the Local and
Sagittarius arms.  At this distance the RCG sequence should begin at
($K_s$, $H$--$K_s$) = (12.5, 0.35), i.e., very close to the limiting
$K_s$ magnitude.  Cooler and absolutely brighter stars of the Local arm
with various reddenings are seen filling the area between $K_s$ = 8.0
and 13.0, with $H$--$K_s$\,$>$\,0.3.

In the $J$--$H$ vs.~$H$--$K_s$ diagram (Figure 5a) the stars with
$J$--$H$\,$<$\,0.6 all belong to the Local arm and are unreddened or
little reddened (up to $E_{H-K_s}$\,$\sim$\,0.1) main-sequence F--G
stars.  The stars lying at the lower edge of the reddening line of
giants can be B-A-F stars of the Local arm considerably reddened by the
Vulpecula cloud and the Vul OB1 association cloud; some of them can be
pre-main-sequence stars (YSOs).  The stars on the giant sequence with
$J$--$H$\,$<$\,0.9 correspond to the Local arm in the solar vicinity and
the interarm region, while at larger $J$--$H$ values most stars belong
to the intersection of the Local and the Sagittarius arms.
Along the reddening line, K and M giants of various spectral subclasses
should be mixed together, and the slope of the line is
2.04\,$\pm$\,0.16 (Table 1).

\subsubsectionb{4.3.2}{$\ell$ = 50\degr, $b$ = 0\degr}

As was discussed in Section 4.1, this direction coincides with the line
of sight along the Sagittarius arm.  In the $J$--$H$ vs.~$H$--$K_s$
diagram (Figure 5b) the reddening line consists mostly of RCGs (and
other stars of the red giant branch) located at various distances and
reddened by different cloud groups.  The RCGs lying lower than $J$--$H$
= 1.0 are affected only by the Local arm clouds.  However, the $K_s$
vs.~$H$--$K_s$ diagram (Figure 2) shows that these stars are located in
the inter-arm space, close to the Sagittarius arm (see discussion in
Section 4.1).  The RCGs with $J$--$H$ between 1.0 and 1.5 are
distributed along the Sagittarius arm.  The stars with
$J$--$H$\,$>$\,1.5 in the CMD all lie above and right of the RCG
sequence and should be field giants of spectral types later than
K3\,III.  In Figure 5b the majority of heavily reddened stars are
scattered between the reddening lines of the slopes $E_{J-H}/E_{H-K_s}$
= 1.9 and 2.1, the mean slope at $J$--$H$ = 2.0 being 1.99\,$\pm$\,0.14.

However, the reddest stars exhibit a tendency to deviate down, to the
lower values of color-excess ratio.  No doubt, this effect is related
mainly to the curvature of the reddening lines due to the band-width
effect (Strai\v zys \& Lazauskait\.e 2008).  Partly, the location of the
reddest stars can be the result of the selection effect caused by
bending down of the intrinsic red-giant sequence for spectral types
cooler than M6\,III (most of them should be long-period variables).  In
Figure 4b we show the $J$--$H$ vs.~$H$--$K_s$ diagram for stars located
in a semi-transparent area of a diameter of 15\arcmin\ at the Galactic
center, in the direction of Baade's Window.  The stars near the end of
the giant branch with $J$--$H$ = 1.0--1.2, when shifted along the
reddening line with the slope $\sim$\,2.0 up to $J$--$H$ = 2.5--3.0,
will appear at the position corresponding to the reddening lines of RCGs
with the slope 1.6--1.8.  A part of these reddest stars can also be
heavily reddened carbon stars since their intrinsic position in the
$J$--$H$ vs.~$H$--$K_s$ diagram is also lower than of M0--M5 III stars.
At the end of the reddening line we observe only the coolest oxygen- and
carbon-rich AGB stars due to a selection effect:  they are seen at
larger distances (and higher reddenings) due to higher luminosity.  At
this distance (and reddening) RCGs apparently are too faint.  Therefore,
in determining the slope of the reddening line we do not use stars
redder than $J$--$H$\,$>$\,2.2.

\subsubsectionb{4.3.3}{$\ell$ = 40\degr, $b$ = 0\degr}

In this direction the line of sight crosses a quite dark region of the
Aquila Rift.  In the CMD the RCG sequence shows that the reddening in
the Local arm is $E_{H-K_s}$\,$\approx$\,0.15 or $A_V$ = 2.5, i.e., it
is larger than at 50\degr.  The Sagittarius arm is crossed twice:
between 2.5--3.5 kpc and at $\sim$\,8 kpc.  The largest extinction (up
to $A_V$\,$\approx$\,8) is observed at the far end of the first
crossing, at 3.0--3.5 kpc.  In the second crossing of the Sagittarius
arm, RCGs are too faint to be visible in $K_s$.  However, M-type giants
and carbon stars can be present in the sample.

The diagram $J$--$H$ vs.~$H$--$K_s$ (Figure 5c) confirms that the
reddening in this direction of the Local arm is larger; the
concentration of red giants begins at $J$--$H$ = 1.1.  The `tail' formed
by reddened stars in the Sagittarius arm is also longer.  The
distribution of stars at $J$--$H$\,=\,2.0 is in accordance with
the mean reddening line slope of 1.98\,$\pm$\,0.16.

\subsubsectionb{4.3.4}{$\ell$ = 30\degr, $b$ = 0\degr}

The area is at the Aquila and Scutum border, in the direction of a
protrusion of the Aquila Rift.  The line of sight, after leaving the
Local arm, at a distance of $\sim$\,2 kpc crosses the Sagittarius arm
and at a distance 3.5 kpc it enters and runs along the Scutum arm.  The
Galactic bulge or bar are not reached yet.  The CMD of this area is
rather complicated, and it is difficult to disentangle the effects of
reddening in the Local, Sagittarius and Scutum arms.  It seems, however,
that the reddening in this direction is larger than at $\ell$ = 40\degr.
In the $J$--$H$ vs.~$H$--$K_s$ diagram (Figure 5d) the mean slope of
the reddening line of red giants at $J$--$H$ = 2.0 is 1.98\,$\pm$\,0.16.

\subsubsectionb{4.3.5}{$\ell$ = 20\degr, $b$ = 0\degr}

The area is in Scutum, at the southern edge of the Aquila Rift.  In this
direction the line of sight crosses the Sagittarius and Scutum arms and
at $\sim$\,5 kpc enters the bulge (including the end of the bar and the
beginning of the Near 3-kpc arm).  In the CMD the RCG sequence runs from
($K_s$, $H$--$K_s$) = (9.0, 0.3) to (12.0, 0.6), i.e., includes RCGs
reddened in the Local, Sagittarius and Scutum arms together.  At the
same time at ($K_s$, $H$--$K_s$) = (8.5, 0.9) the tip of the sequence of
the bulge giants appears.  At (12.0, 0.6) it converges with the RCG
sequence.

In the $J$--$H$ vs.~$H$--$K_s$ diagram (Figure 5e) red giants cover the
reddening line without jumps, with the largest concentration at $J$--$H$
= 1.5.  The stars with redder colors probably all belong to the bulge
and are M-type giants.  The slope of the reddening line at $J$--$H$ =
2.0 is 1.98\,$\pm$\,0.15.

\subsubsectionb{4.3.6}{$\ell$ = 10\degr, $b$ = 0\degr}

The area is in Sagittarius, between the Small Sagittarius Cloud (M\,24)
and the Trifid Nebula (M\,20).  In this direction the line of sight
crosses the Local, Sagittarius and Scutum arms and enters deep into the
bulge.  In the CMD the RCG sequence is not sufficiently definite, but a
dense clump at ($K_s$, $H$--$K_s$) = (11.0, 0.4) is seen.  This clump
has an extension upward which probably is the red giant sequence
reddened by the Local and Sagittarius arms.  Also, the area of the bulge
M-type giants with larger reddening is seen, but not so concentrated as
in the $\ell$ = 20\degr\ area.

The $J$--$H$ vs.~$H$--$K_s$ diagram of the area (Figure 5f) is similar
to that at $\ell$ = 20\degr, but the reddening line is much richer at
large reddenings.  The mean slope of the reddening line at $J$--$H$ =
2.0 is 1.94\,$\pm$\,0.14.

\subsubsectionb{4.3.7}{$\ell$ = 0\degr, $b$ = 0\degr}

This is the area in the direction to the Galactic center.  The structure
of the CMD in this direction is similar to that for $\ell$ = 2.5\degr\
described in Section 4.2 but with some important differences.  The
number of stars everywhere is much smaller, probably due to large
interstellar extinction.  The RCG sequence is seen only down to ($K_s$,
$H$--$K_s$) = (11.0, 0.3).  The RCGs in Sagittarius and Scutum arms
between $K_s$ = 11--13 are absent; probably, they are reddened so
strongly that have been moved down along reddening lines lower than the
limiting $K_s$ magnitude.  However, the tip of the sequence of bulge
M-type giants is well seen at ($K_s$, $H$--$K_s$) = (9.5, 1.4).

The $J$--$H$ vs.~$H$--$K_s$ diagram of the area (Figure 5g) is also
peculiar.  The unreddened or little reddened red giants of the Local arm
and, maybe, of the beginning of the Sagittarius arm, are seen up to
$J$--$H$\,$\approx$\,1.3.  The bulge M-giants are strongly shifted along
the reddening line and their density increases considerably only at
$J$--$H$\,$>$\,2.5, reaching maximum at about 3.0.  The mean reddening
of the upper tip of the bulge giant sequence estimated from the CMD is
$E_{H-K_s}$\,$\approx$\,1.1, this corresponds to $A_V$ = 18.7 mag.
This is the mean value of the extinction over an area of 1\degr\
diameter.  Some individual stars exhibit $H$--$K_s$\,$\geq$\,2.0--2.5,
and this corresponds to $A_V$\,$\approx$\,40 mag.  This is in
agreement with the results of Schultheis et al.  (1999) using the DENIS
survey data and Cotera et al.  (2000) using the AAS {\it JHK}\arcmin\
system.

The mean slope of the reddening line of red giants at $J$--$H$ = 2.0 in
the direction of the Galactic center does not differ from other areas
discussed above, its value is 2.02\,$\pm$\,0.18.  The stars with
$J$--$H$\,$>$\,3.0 deviate down; part of them can be the AGB stars of
the bulge (see the discussion in subsection 4.3.2).

\subsubsectionb{4.3.8}{$\ell$ = 350\degr, $b$ = 0\degr}

The area is in Scorpius, its line of sight crosses the Local,
Sagittarius, Scutum, Norma, 3-kpc arms and the bulge.  Although
the area is symmetrical with respect to the Galactic center to the
$\ell$ = 10\degr\ area, their CMDs differ considerably.  At 350\degr\ we
see quite distinct RCG sequence consisting of two parts, corresponding
probably to the Sagittarius and the Scutum arms:  it runs from ($K_s$,
$H$--$K_s$) = (9.0, 0.25) to (11.0, 0.6).  The upper part of the bulge
giant sequence at ($K_s$, $H$--$K_s$) = (9.5, 1.0) is also more
populated.

The $J$--$H$ vs.~$H$--$K_s$ diagram of the area (Figure 5h) is very
similar to that for \hbox{$\ell$ =} 10\degr.  The only significant
difference is that at $\ell$ = 350\degr\ the maximum reddening is
smaller.  The mean slope of the reddening line of red giants at $J$--$H$
= 2.0 is 1.96\,$\pm$\,0.15.

\subsubsectionb{4.3.9}{$\ell$ = 340\degr, $b$ = 0\degr}

The area is at the Scorpius, Ara and Norma corner.  In this direction
the line of sight crosses the Local, Sagittarius, Scutum, Norma and
3-kpc arms.  However, the distance of the Norma arm is at 5 kpc,
i.e., its RCGs are fainter than the limiting magnitude. The CMD of
this area, in comparison to the mirror area at $\ell$ = 20\degr, shows
the same features, but the RCG sequence corresponding to the Sagittarius
and Scutum arms at $\ell$ = 340\degr\ is much more populated.  This
effect could be of instrumental origin if in the southern sky 2MASS
magnitudes were measured with higher accuracy.  More populated are and
other areas of the Southern sky.

The $J$--$H$ vs.~$H$--$K_s$ diagram of the area (Figure 5i) is more
abundant in red giants at low reddenings than the diagram for $\ell$ =
20\degr. Otherwise, both diagrams are similar, exposing the same slope.

\subsubsectionb{4.3.10}{$\ell$ = 330\degr, $b$ = 0\degr}

The area is projected on one of the dark lanes in Norma.  In this
direction the line of sight crosses the Local, Sagittarius and Scutum
arms and then runs along the Norma arm.  In comparison with the area at
$\ell$ = 30\degr\ on the other side of the Galactic center, its RCG
branch in CMD is much more populated and is bluer:  it runs from ($K_s$,
$H$--$K_s$) = (9.0, 0.3) to (12.0, 0.6).  This is seen in the rich
$J$--$H$ vs.~$H$--$K_s$ diagram (Figure 5j) where high concentration of
red giants starts at $J$--$H$ = 0.75.  All this shows that the
extinction in the Norma clouds is much lower than in the Aquila Rift
clouds.  However, the slope of the reddening line is similar in both
directions.

\subsubsectionb{4.3.11}{$\ell$ = 320\degr, $b$ = 0\degr}

This area is located in the direction of the same Norma dark lane, only
4\degr\ from $\alpha$ Cen.  The line of sight crosses the Sagittarius
and Scutum arms (the last arm twice).  CMDs of this area and the mirror
area at $\ell$ = 40\degr\ are quite similar, but the area of the
southern sky is again richer is stars.  The reddening in the 320\degr\
area is also smaller than in its northern counterpart:  the RCG sequence
runs from ($K_s$, $H$--$K_s$) = (9.0, 0.25) to (12.5, 0.6).  The
$J$--$H$ vs.~$H$--$K_s$ diagrams (Figures 4c and 4k) of both areas are
similar, but in the southern area the density of red giants is larger at
lower reddenings.  The slope of the reddening line at $J$--$H$ = 2.0 is
1.99\,$\pm$\,0.14.

\subsubsectionb{4.3.12}{$\ell$ = 310\degr, $b$ = 0\degr}

This area is located in a relatively transparent direction between the
star $\alpha$ Cen and the Coalsack.  In this direction, after crossing
the Local and Sagittarius arms, the line of sight runs along the Scutum
arm.  CMD and $J$--$H$ vs.~$H$--$K_s$ diagrams for this area (Figure
5$\ell$) are quite similar to the diagrams for its mirror area in the
northern sky at $\ell$ = 50\degr\ (Figures 2a and 5b), however, the
310\degr\ area contains many more stars, as the other southern areas.
In the CMD the RCG sequence extends from ($K_s$, $H$--$K_s$) = (9.0,
0.2) to (12.3, 0.65), being much broader than at 50\degr.  The slope of
the reddening line is 2.00\,$\pm$\,0.14.

%%%%%%%%%%%%%%%%%%%%%%%%%%%%%%%%%%%% TABLE 2

\begin{table}[!t]
\begin{center}
\vbox{\small\tabcolsep=6pt
\parbox[c]{124mm}{\baselineskip=10pt
{\normbf\ \ Table 2.}{\norm\
The investigated areas located in star-forming areas.  The
given mean ratios of color excesses correspond to $J$--$H$ between
1.6--2.2 for most areas, but between 1.2--2.0 in the Lup~1 cloud.  The
last column shows the number of stars used to calculate the ratio of
color excesses.
\lstrut}}
\begin{tabular}{llrrcr}
\tablerule
Area &  Number in Dobashi &
\multicolumn{2}{c}{Center}&
$E_{J-H}/E_{H-K_s}$ & Number \\
  &    et al. (2005)      & $\ell$\degr\ & $b$\degr\  &    & of stars \\
\tablerule
Ser core       & T\,289-P12   &  31.6 &  +5.3 & $2.096\pm 0.215$ &   62~~~ \\[-1pt]
Ser cloud      & T\,279-P7    &  28.8 &  +3.5 & $1.983\pm 0.165$ &  502~~~ \\[-1pt]
Ser cloud      & T\,243-P1/2  &  23.5 &  +8.3 & $2.052\pm 0.188$ &   65~~~ \\[-1pt]
Pipe nebula    & T\,25-P1     &   1.6 &  +3.4 & $1.912\pm 0.206$ &  908~~~ \\[-1pt]
CrA cloud      & T\,2213-P1/2 & 359.9 & --17.8 & $1.787\pm 0.160$ &    9~~~ \\[-1pt]
Pipe nebula    & Barnard 59   & 357.2 &  +7.1 & $1.925\pm 0.176$ &   92~~~ \\[-1pt]
Oph cloud lane & T\,2171-P1,  & 354.0 & +15.8 & $1.934\pm 0.135$ &   65~~~ \\[-1pt]
Oph cloud lane & T\,2171-P5,  & 354.9 & +15.0 &  --              &   --~~~~ \\[-1pt]
Rho Oph cloud  & T\,2171-P3   & 353.0 & +17.0 & $1.915\pm 0.177$ &   50~~~ \\[-1pt]
Lup1 cloud     & T\,2072      & 338.0 & +16.4 & $2.087\pm 0.221$ &   18~~~ \\[-1pt]
Lup1 cloud     & T\,2079-P1/2 & 338.8 & +17.4 &  --              &   --~~~~ \\[-1pt]
Lup3 cloud     & T\,2084      & 339.5 &  +9.3 & $1.900\pm 0.165$ &   19~~~ \\
\tablerule
\end{tabular}
}
\end{center}
\vskip-6mm
\end{table}

\newpage

\subsectionb{4.4}{Two-color diagrams for the Gould Belt areas
containing dense dust clouds}

In Figure 6 (a--j) we present the $J$--$H$ vs.~$H$--$K$ diagrams for
some areas of the Local arm located towards the inner Galaxy (Table 2).
Most of them are known star-forming regions within the Gould Belt.  The
slopes of the reddening lines of red giants were calculated for the
stars with $J$--$H$ between 1.6 and 2.2.  In the Lupus areas this
interval was 1.2--2.0.  As in the case of the Galactic plane areas, we
have excluded possible AGB stars and YSOs located below the reddening
line of red giants with the slope 1.6.

\subsubsectionb{4.4.1}{The Serpens dust/molecular cloud}

In our investigation the direction to the Serpens section of the Aquila
Rift is represented by three areas, coinciding with the dust clumps
T\,289-P12, T\,279-P7 and T\,243-P1/2 in the Dobashi et al.  (2005)
atlas.  The $J$--$H$ vs.~$H$--$K_s$ diagrams for stars in these three
areas are plotted in Figures 6 (a), (b) and (c).  The Serpens clouds are
at a distance of about 225 pc (Strai\v{z}ys et al. 1996, 2003).  Most
probably, in the direction of the first and the third areas this is the
only dust layer in the line of sight.  In this case the distribution of
distant red clump giants and of cooler K and M giants along the
reddening line is caused by a differential reddening originating in this
single layer.  The second Serpens area is at lower Galactic latitude
($b$ = +3.5\degr), and the line of sight can cross more dust clouds at
larger distances.  This area has the largest surface density of dust
with the maximum extinction $A_V$ of 28 mag, while in the first and the
third area the extinction is about 20 and 15 mag, respectively.  In all
diagrams the stars lying lower than the reddening line of red giants are
observed.  Their possible origin was discussed in Section 3.
%\enlargethispage{5mm}

In all three diagrams we observe the rise of density of red giants at
$J$--$H$\,$\approx$\,1.0, which corresponds to $A_V$ = 3.4.  Although
the lengths of reddening lines in the three directions are quite
different, their slopes are similar:  the stars are
scattered between the color-excess ratios of 1.8 and 2.1, the mean
values of slopes being 2.10\,$\pm$\,0.21, 1.98\,$\pm$\,0.16 and
2.05\,$\pm$\,0.19 (Table 2).

%%%%%%%%%%%%%%%%%%%%%%%%%%%%%%%%  FIGURES 6

\vbox{
\centerline{\psfig{figure=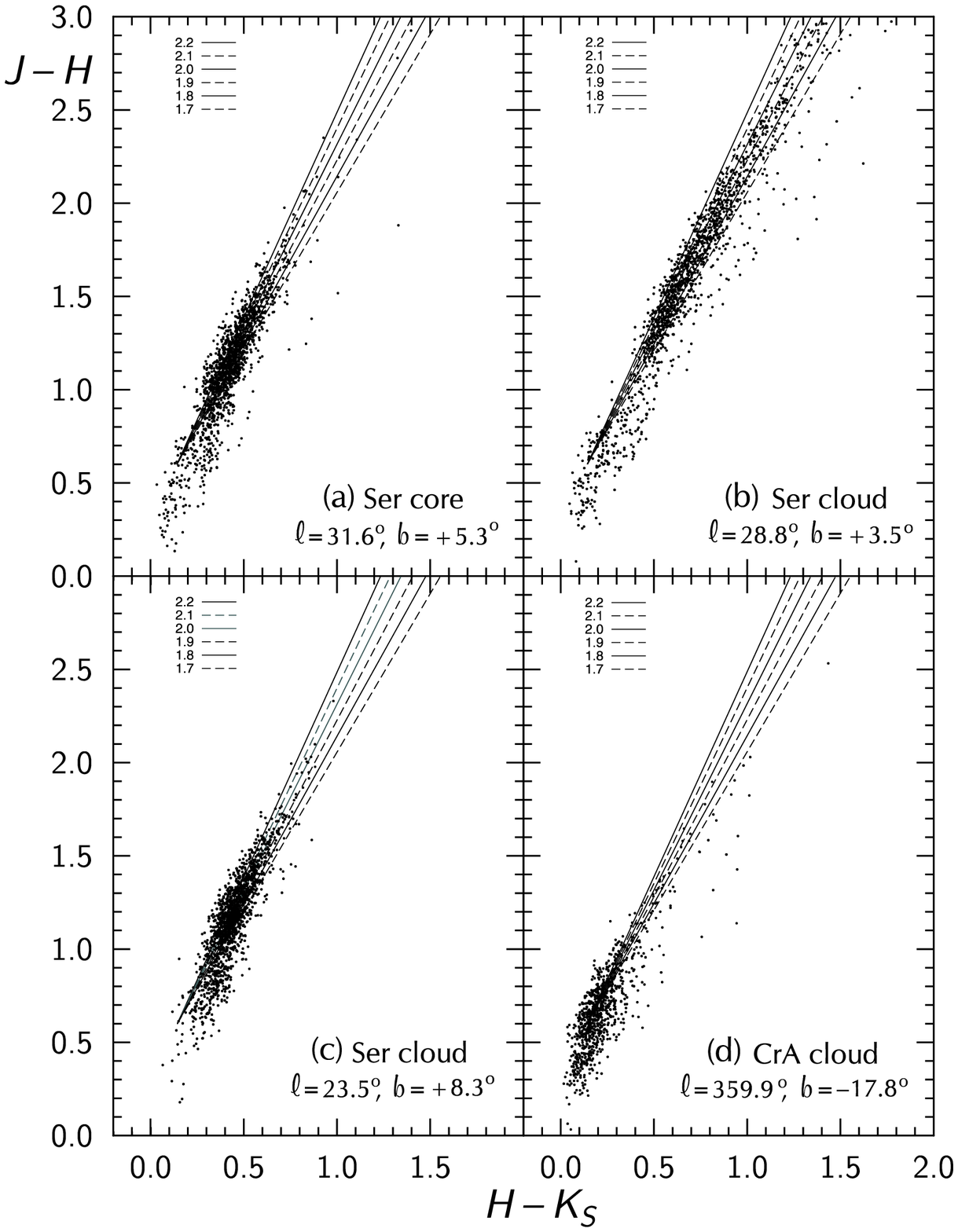,angle=0,width=120truemm,clip=}}
\vskip2mm
\captionb{6\,(a--d)}{Two-color diagrams for the star-forming areas: (a) --
the Serpens cloud core, (b) and (c) -- two Serpens cloud areas and (d)
-- the CrA cloud.  The six straight lines are the reddening lines of
RCGs with different slopes, explained in the inserts.}
}
\newpage

\vbox{
\centerline{\psfig{figure=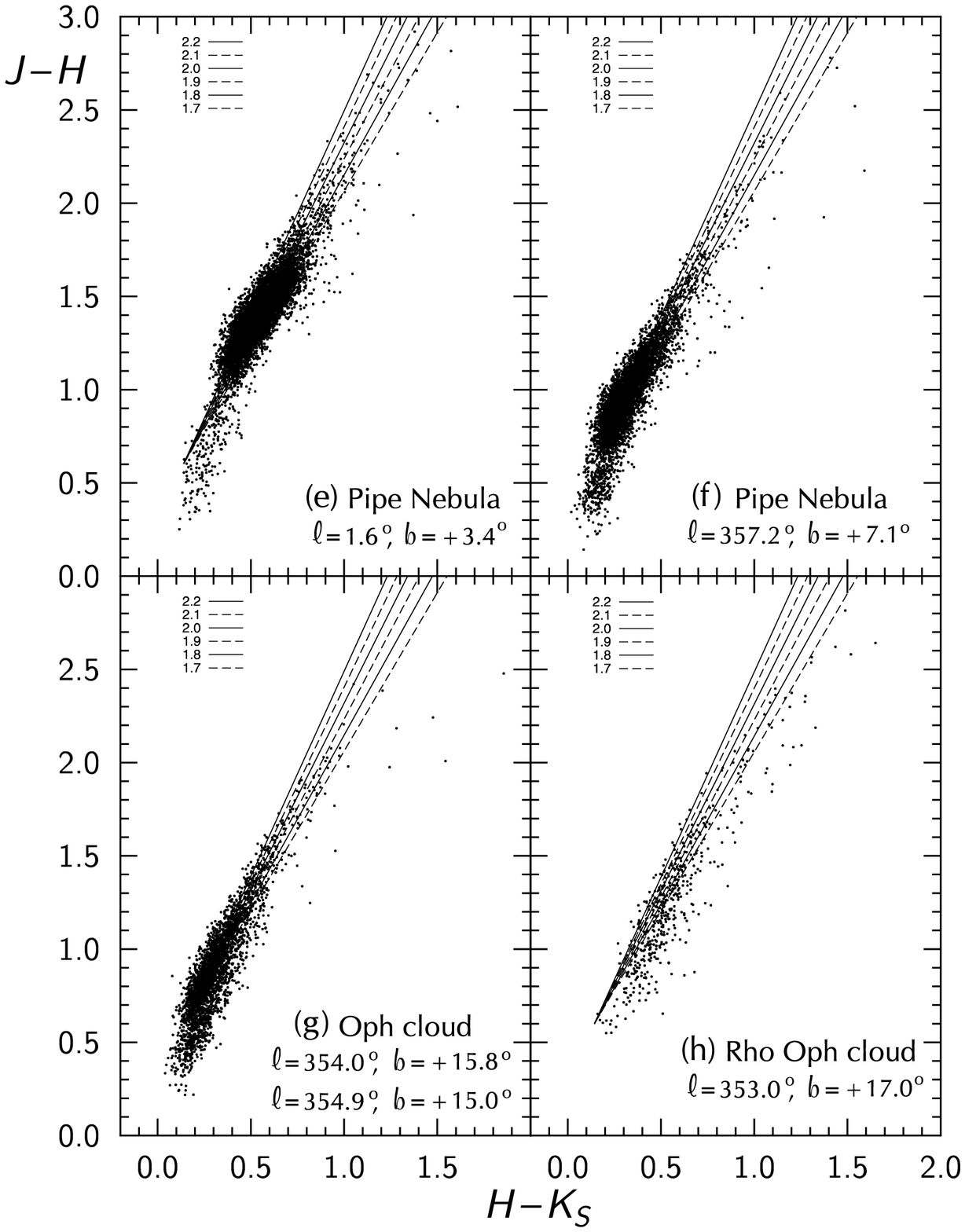,angle=0,width=120truemm,clip=}}
\vskip2mm
\captionb{6\,(e--h)}{The same as in Fig. 6 (a--d) but for the areas: (e) and
(f) -- two areas in the Pipe Nebula, (g) -- two areas in a dark lane of
the Rho Oph cloud and (h) -- the central Rho Oph cloud area.}
}
\newpage

\vbox{
\centerline{\psfig{figure=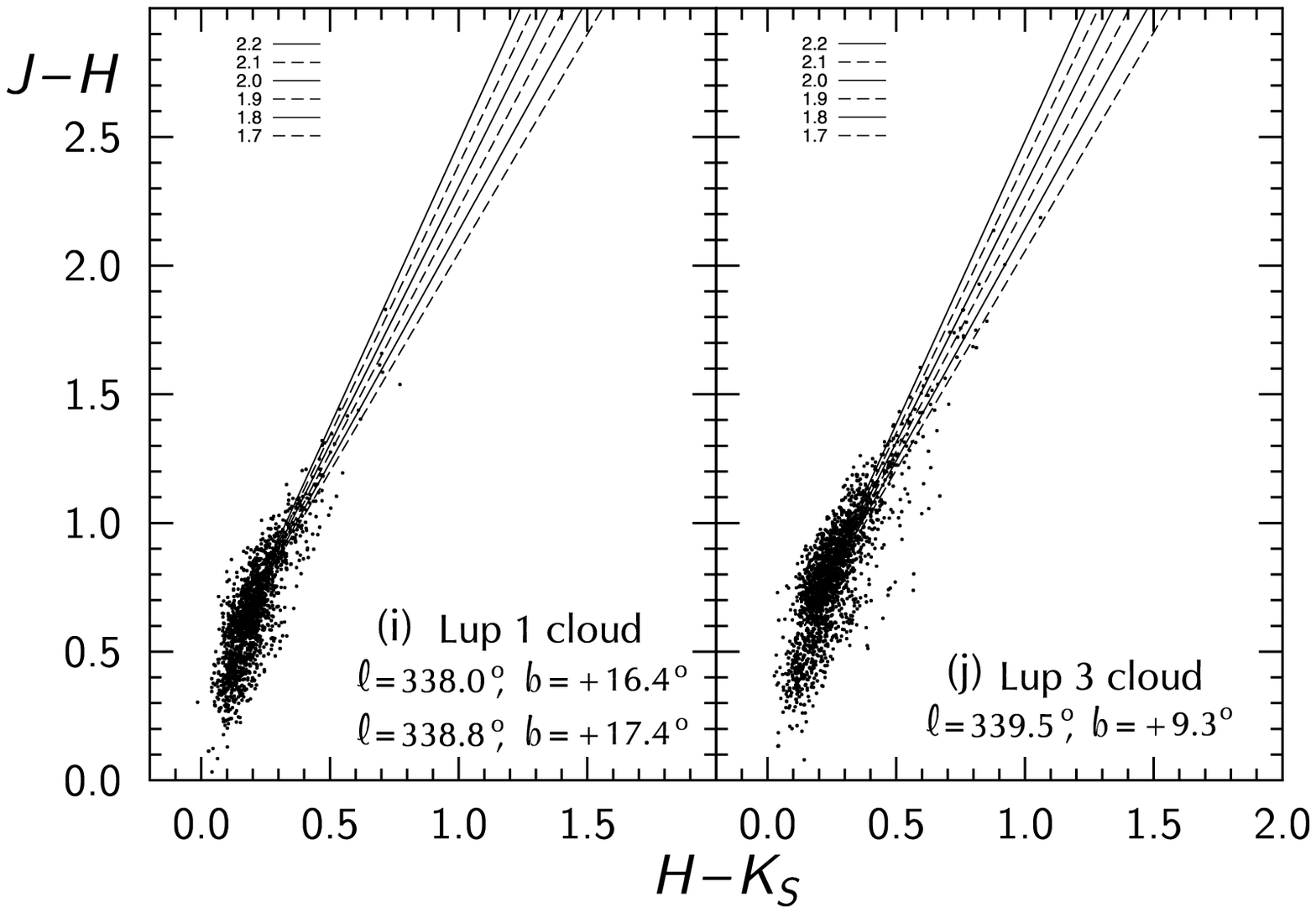,angle=0,width=120truemm,clip=}}
\vskip2mm
\captionb{6\,(i--j)}{The same as in Fig. 6 (a--d) but for the areas: (i) --
two areas in the Lupus 1 cloud and (j) -- an area in the Lupus 3 cloud.}
}

\subsubsectionb{4.4.2}{The Corona Australis dust/molecular cloud}

This cloud with active star formation is represented by the area in its
densest place with the center at $\ell$ = 359.9\degr, $b$ = --17.8\degr.
The most recent review of the region by Neuh\"auser \& Forbrich (2008)
places the cloud at a distance of 130 pc.  Its reddening line of red
giants in the $J$--$H$ vs.~$H$--$K_s$ diagram (Figure 6d) has its end
probably at $J$--$H$\,$\approx$\,2.0; this corresponds to $A_V$ = 11.9
mag.  The reddening line of red giants from below has a well populated
sequence of pre-main-sequence stars.  Although it is difficult to find
the boundary of the two sequences, between $J$--$H$ = 1.0 and 1.6 the
reddening line of red giants is sufficiently well defined.  The mean
slope of 9 stars falling between $J$--$H$ = 1.6 and 2.0 is
1.79\,$\pm$\,0.16, i.e., it is somewhat smaller than in most of other
areas described in Sections 4.3 and 4.4.

\subsubsectionb{4.4.3}{The Pipe Nebula}

The Pipe Nebula dark cloud in the direction of the Galactic center is
located at a distance of about 130 pc and belongs to the
Scorpio-Centaurus subsystem of the Gould Belt (see the review by Alves
et al. 2008).  This dark cloud is represented by two areas at Galactic
latitudes +3.4\degr\ and +7.1\degr.  Since both areas are out of
Galactic plane, the giants located at various distances and seen in the
background all are reddened only by the Pipe Nebula cloud.  As a result,
CMDs in this direction are quite similar to the CMD for Baade's Window
shown in Figure 4a:  in the diagrams the giants located at various
distances form a broad belt extending from $K_s$\,$\approx$\,8.0 to
13.0.  However, the reddening in the Pipe areas is larger than in
Baade's Window.  In the first and the second Pipe areas the lower end of
the belt is at ($K_s$, $H$--$K_s$) = (13.0, 0.45) and (13.0, 0.25),
respectively.

In the $J$--$H$ vs.~$H$--$K_s$ diagram (Figures 6e and 6f) the giant
sequences representing the Pipe Nebula are very rich but most stars are
lower than $J$--$H$ = 1.5 or 2.0, respectively.  At large reddenings
only a few stars are present.  The mean slopes of the reddening lines
at $J$--$H$ = 2.0 in the two areas are 1.91\,$\pm$\,0.21 and
1.92\,$\pm$\,0.18, respectively.

\subsubsectionb{4.4.4}{The Rho Ophiuchi and related clouds}

The Rho Ophiuchi star-forming region, located at a distance of 120--140
pc (Wilking et al. 2008) in our sample is represented by three areas:
one of them is placed on the densest part of the Rho Oph cloud and the
remaining are located on the dust lane extending in the direction of the
bulge.  Since in the last two areas the density of heavily reddened
stars is rather low, in the $J$--$H$ vs.~$H$--$K_s$ diagram (Figure 6g)
we plotted stars from both these areas together.  The diagram for the
Rho Oph area is plotted separately (Figure 6h).  Red giants with
$J$--$H$ between 1.6 and 2.2 in both diagrams follow the reddening lines
with the mean slopes 1.93\,$\pm$\,0.14 and 1.92\,$\pm$\,0.18.

\subsubsectionb{4.4.5}{The Lupus clouds}

For the investigation of the Lupus cloud area we selected three dust
clumps:  two of them are in the Lupus 1 cloud and one is in the Lupus 3
cloud.  According to Comer\'on (2008), the Lupus 1 cloud is located at a
distance of $\sim$\,150 pc and the Lupus 3 cloud at $\sim$\,200 pc.  To
increase the number of heavily reddened stars, the stars from both areas
of the Lupus 1 cloud were plotted on the same graph (Figure 6i).  Even
in this case the most reddened stars are lower than in other areas.  The
slope of reddening line derived from 18 stars between $J$--$H$ = 1.2 and
1.8 is 2.09\,$\pm$\,0.22.  In the Lupus 3 cloud (Figure 6j) the maximum
reddening is much larger.  However, here is a clear dependence of the
slope on the minimum value of $J$--$H$.  Taking 19 stars between
$J$--$H$ = 1.6 and 2.2 we obtain the slope 1.90\,$\pm$\,0.16.  However,
taking 70 stars between $J$--$H$ = 1.25 and 2.2 the slope is
2.00\,$\pm$\,0.25.  Probably, at the largest reddenings we encounter
the influence of YSOs.

\sectionb{5}{DISCUSSION}

In this section we compare our results with the reddening line
slopes obtained in the same range of Galactic coordinates in other
investigations.  To avoid systematic errors, we will take into account
only those studies which are based on the $J$,$H$,$K_s$ photometric data
taken from the 2MASS survey or obtained by special observations in
similar photometric systems.

To our knowledge, no systematic investigations of reddening effects in
the 2MASS $J$--$H$ vs.~$H$--$K_s$ diagram along the Milky Way have been
undertaken so far.  However, star-forming regions of the Local arm, seen
in the direction of inner Galaxy, were investigated in many papers.

Let us start from the direction toward the Galactic bulge.  Recently,
two investigations of the reddening law in infrared were published by
Nishiyama et al.  (2006, 2008).  The authors have used deep {\it J, H},
$K_s$ photometry in the SIRIUS photometric system obtaining
$E_{J-H}/E_{H-K_s}$ = 1.72 and 1.74.  However, the SIRIUS system is
different from the 2MASS system (Kato et al. 2007; Ku\v{c}inskas et al.
2008).  After reduction of the Nishiyama et al. colors to the 2MASS
system, we obtain the color-excess ratio close to 2.0.

In two other papers 2MASS data have been used to investigate the
reddening in the direction of the Pipe Nebula, including the dark cloud
B\,59.  Lombardi et al.  (2006) in a large area of
8\degr\,$\times$\,6\degr\ find that the ratio $E_{J-H} / E_{H-K_s} =
1.85 \pm 0.15$.  Brooke et al.  (2007) in their {\it Spitzer} project
``Cores to Disks'' for the B\,59 cloud show the $J$--$H$ vs.~$H$--$K_s$
diagram with the reddening line having a slope of 1.92.  An extremely
low value of the slope, 1.36--1.52, has been obtained by
Rom\'an-Z\'uniga et al.  (2007) in their investigation of the extinction
law in the core of the B\,59 cloud.  Most probably, this result can be
explained by the ignorance of the band-width effect, i.e., the curvature
of the reddening line. As it is shown by Strai\v zys \& Lazauskait\.e
(2008), this effect is quite strong at heavy reddenings.

%%%%%%%%%%%%%%%%%%%%%%%%%%%%%%%%  FIGURE 7

\begin{figure}[!t]
\vbox{
\centerline{\psfig{figure=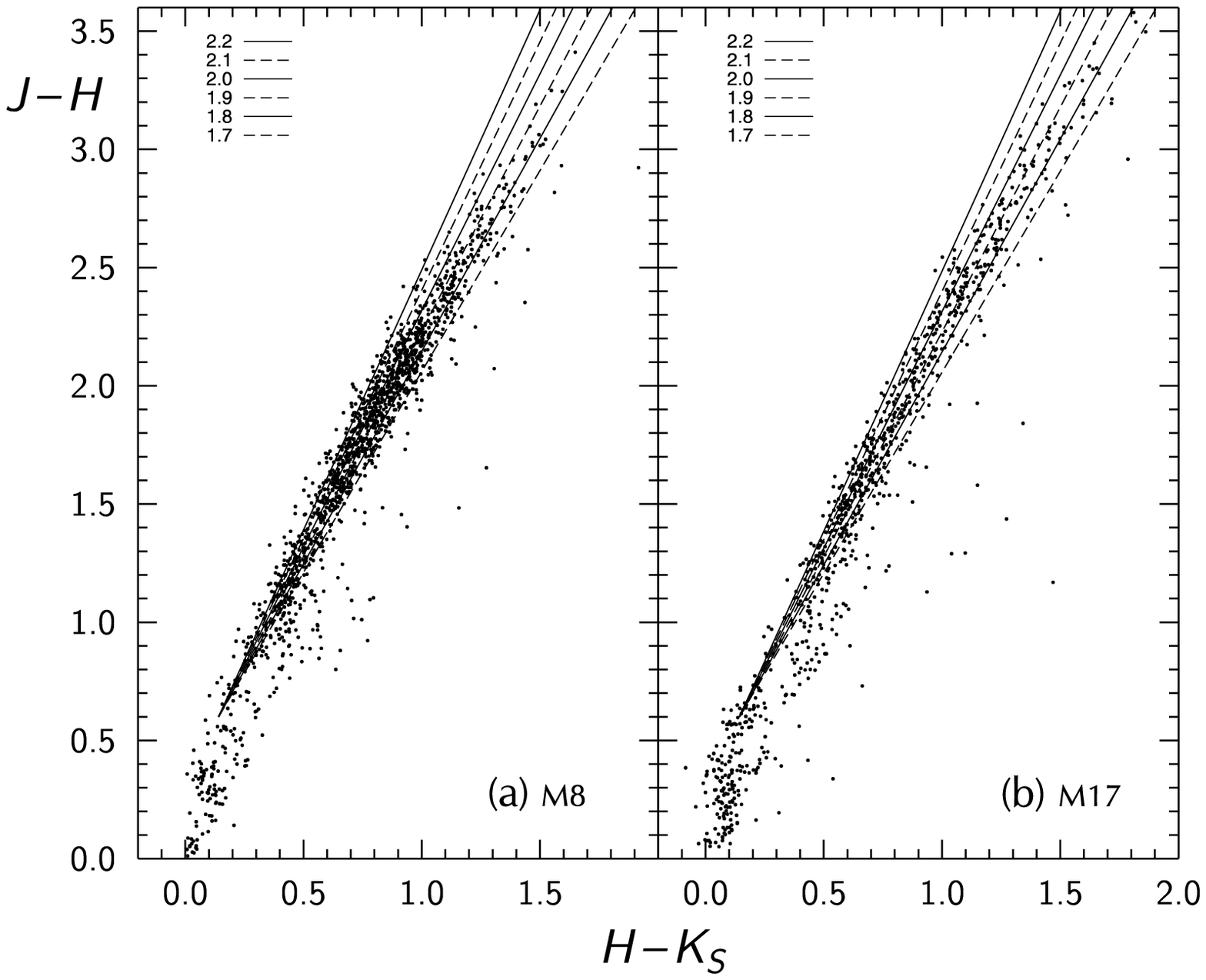,angle=0,width=120truemm,clip=}}
\vskip2mm
\captionb{7}{The same as in Figure 6 but for the areas:  (a) -- M\,8 (Lagoon
Nebula) and (b) -- M\,17 (Omega Nebula).}
}
\end{figure}

Arias et al.  (2006) in the {\it J,H},$K_s$ system investigated the
Hourglass Nebula within the Lagoon Nebula (M\,8) at $\ell$ = 6.0\degr,
$b$ = --1.2\degr\ from the images obtained at the Las Campanas
Observatory.  Although the authors accept a slope of the reddening line
from Rieke \& Lebofsky (1985), the distribution of the reddened field
stars in the $J$--$H$ vs.~$H$--$K_s$ diagrams for the Hourglass Nebula
and the nearby control field are consistent with a higher slope of
1.9--2.0.  In Figure 7a we show the $J$--$H$ vs.~$H$--$K_s$ diagram
plotted for an area with a diameter of 0.5\degr\ in the direction of
M\,8 for stars with the accuracy of 2MASS magnitudes of $\leq$\,0.03
mag.  The mean ratio of color excesses for 332 stars with $J$--$H$
between 1.8 and 2.2 is 1.97\,$\pm$\,0.16.

Two {\it J, H}, $K_s$ investigations in the M\,17 region (Omega Nebula)
at $\ell$ = 15\degr, $b$ = --0.7\degr\ are available.  Jiang et al.
(2002) using the SIRIUS system obtain the reddening line slope of 1.9.
Hoffmeister et al.  (2008) from imaging observations on VLT in another
area of M\,17 obtain the ratio 2.06.  Unfortunately, photometric systems
of both investigations are not sufficiently described, so we are not
sure if these ratios are valid for the 2MASS system.  In Figure 7b we
plot the stars in the direction of M\,17 taking 2MASS magnitudes with an
accuracy of $\leq$\,0.03 mag. The mean ratio of color excesses for 111
stars with $J$--$H$ between 1.8 and 2.2 is 1.96\,$\pm$\,0.15.

The star-forming complex Rho Ophiuchi was long known as the region with
a low ratio of $E_{J-H} / E_{H-K}$ (1.57, Kenyon et al. 1998;
1.60--1.68, Naoi et al. 2006).  The determination of the RCG reddening
line slope in this region is complicated by abundant population of YSOs
which in the $J$--$H$ vs.~$H$--$K_s$ diagram lie lower than red giants.
However, YSOs are less abundant in the dark lanes extending from the
central Rho Ophiuchi cloud.  Our Figure 6h representing the clumps P1
and P5 of the T\,2171 cloud shows that at $J$--$H$ = 2.0 the slope of
the RCG line is consistent with values between 1.7 and 2.1, the
average slope being 1.93\,$\pm$\,0.14.  A similar slope can be derived
from Fig.\,5 of Lombardi et al.  (2008) plotted from the 2MASS data
(including the observations with lower accuracy) for a large region
covering Ophiuchus and Lupus.

Haas et al. (2008) plotted the $J$--$H$ vs.~$H$--$K_s$ diagram for the
R Coronae Australis cloud stars imaged with the CTIO 4 m Blanco
telescope and obtained a slope of 1.8. This value is consistent with
our diagram shown in Figure 6d, if we exclude possible pre-main-sequence
objects. In Table 2 we give a value of 1.79\,$\pm$\,0.16.

Indebetouw et al.  (2005) included the region at $\ell$ = 42\degr, $b$ =
+0.5\degr\ in Aquila in the investigation of the infrared interstellar
extinction law.  Their value of $E_{J-H} / E_{H-K_s}$ = 1.78 based on
the 2MASS data is not very different from our estimate at $\ell$ =
40\degr\ (1.98\,$\pm$\,0.16, Figure 5c).

\sectionb{6}{CONCLUSIONS}

The results of this investigation allow us to conclude that in the
direction of the inner Galaxy, between the longitudes 60\degr\ and
310\degr, the ratio $E_{J-H} / E_{H-K_s}$ of red giants shows quite
small variations.  The value of the ratio measured for red giants at
$J$--$H$\,$\approx$\,2.0 is between 1.9 and 2.0 both in the clouds of
the Local arm and in the Sagittarius arm.  The Galactic bulge red giants
reddened by interstellar clouds of the Local, Sagittarius, Scutum, Norma
and 3-kpc arms also follow the same reddening law.  In the 2MASS
system the `normal' value of the ratio probably is close to 1.9--2.0,
i.e., it is larger by 0.2--0.3 than it was usually considered
modeling the interstellar dust.

Most values of $E_{J-H} / E_{H-K_s}$ reported till now in the
literature, except for a few exclusions, are between 1.6--1.8, i.e.,
they are lower than the values obtained above.  We conclude that the
reduction of the observed ratios can be conditioned by several reasons:
(1) the ignorance of the fact that interstellar reddening lines in the
$J$--$H$ vs.~$H$--$K_s$ diagrams are not straight but due to the
band-width effect at large reddenings they bend down; (2) in the areas
near the Galactic equator and towards the bulge most luminous and most
reddened stars are oxygen- or carbon-rich AGB stars whose intrinsic
positions in the $J$--$H$ vs.~$H$--$K_s$ diagram bend down from the
intrinsic sequence of K--M5 giants; (3) in some star-forming areas the
slope of the reddening line can be reduced by unrecognized young stellar
objects (YSOs) in various stages of evolution; many of them are
surrounded by warm dust envelopes and disks which place them in the
$J$--$H$ vs.~$H$--$K_s$ diagram lower than the reddening line of RCGs.

One more source of disagreement between different determinations of
color-excess ratios in the same area is related to the variety of
photometric systems used, even if they all have the same designation
({\it J, H}, $K_s$).  The observers in some cases use only a few
standard stars with their magnitudes imported from other (similar)
photometric system, without any color corrections.  Other sources of
significant systematic errors can be related to insufficient atmospheric
extinction corrections (average extinction coefficients accepted),
response non-linearity of the detectors, inhomogeneities in the filter
transmittance curves, etc. All these effects should be carefully
taken into account.

\thanks{We are grateful to Edmundas Mei\v{s}tas for his help preparing
the figures. The use of the 2MASS, SkyView and Simbad databases is
acknowledged.}

\References

\refb Alves D. R. 2000, ApJ, 539, 732

\refb Alves J., Lombardi M., Lada Ch.  J. 2008, in {\it Handbook of Star
Forming Regions}, vol.\,2, ed.  B. Reipurth, ASP, p.\,415

\refb Arias J. I., Barb\'a R. H., Ma\'iz Appell\'aniz J. et al. 2006,
MNRAS, 366, 739

\refb Benjamin R. A. et al. 2008, BAAS, 40, 266; NASA News, June 3, 2008

\refb Brooke T. Y., Huard T. L., Bourke T. L. et al. 2007, ApJ, 655, 364

\refb Comer\'on F. 2008, in {\it Handbook of Star Forming Regions},
vol.\,2, ed.  B. Reipurth, ASP, p.\,295

\refb Cotera A. S., Simpson J. E., Erickson E. F. et al. 2000, ApJS,
129, 123

\refb Cutri R. M., Skrutskie M. F., Van Dyk S., Beichman C. A. et al.
2006, Eplanatory Supplement to the 2MASS All Sky Data Release and
Extended Mission Products, \\
http://www.ipac.caltech.edu/2mass/releases/allsky/doc/explsup.html

\refb Dutra C. M., Santiago B. X., Bica E. 2002, A\&A, 381, 219

\refb Dutra C. M., Santiago B. X., Bica E.\,L\,D., Barbuy B. 2003,
MNRAS, 338, 253

\refb Eiroa C., Djupvik A. A., Casali M. M. 2008, in {\it Handbook of
Star Forming Regions}, vol.\,2, ed.  B. Reipurth, ASP, p.\,693

\refb Forbes D. 1985, AJ, 90, 301

\refb Frogel J. A., Tiede G. P., Kuchinski L. E. 1999, AJ, 117, 2296

\refb Grocholski A. J., Sarajedini A. 2002, AJ, 123, 1603

\refb Groenewegen M.\,A.\,T. 2008, A\&A, 488, 935

\refb Haas M., Heymann F., Domke I. et al. 2008, A\&A, 488, 987

\refb Hoffmeister V. H., Chini R., Scheyda C. M. et al. 2008, ApJ, 686,
310

\refb Indebetouw R., Mathis J. S., Babler B. L. et al. 2005, ApJ, 619,
931

\refb Jiang Z., Yao Y., Yang J. et al. 2002, ApJ, 577, 245

\refb Kato D., Nagashima C., Nagayama T. et al. 2007, PASJ, 59, 615

\refb Kenyon S. J., Lada E. A., Barsony M. 1998, AJ, 115, 252

\refb Ku\v{c}inskas A., Dobrovolskas V., Lazauskaite R., Lindegren L.,
Tanabe T. 2008, Baltic Astronomy, 17, 283 (this issue)

\refb Lombardi M., Alves J., Lada C. J. 2006, A\&A, 454, 781

\refb Lombardi M., Lada C. J., Alves J. 2008, A\&A, 489, 143

\refb Naoi T., Tamura M., Nakajima Y. et al. 2006, ApJ, 640, 373

\refb Neckel Th., Klare G. 1980, A\&AS, 42, 251 %%% (Vul cloud)

\refb Neuh\"auser R., Forbrich J. 2008, in {\it Handbook of Star Forming
Regions}, vol.\,2, ed.  B. Reipurth, ASP, p.\,735

\refb Nishiyama S., Nagata T., Kusakabe N. et al. 2006, ApJ, 638, 839

\refb Nishiyama S., Nagata T., Tamura M. et al. 2008, ApJ, 680, 1174

%\refb Persson S. E., Murphy D. C., Krzeminski W. et al. 1998, AJ, 116,
%2475

\refb Prato L., Rice E. L., Dame T. M. 2008, in {\it Handbook of Star
Forming Regions}, vol.\,1, ed.  B. Reipurth, ASP, p.\,18

\refb Racca G., G\'omez M., Kenyon S. J. 2002, AJ, 124, 2178

\refb Rieke G. H., Lebofsky M. J. 1985, ApJ,  288, 618

\refb Rom\'an-Z\'uniga C. G., Lada C. J., Muench A., Alves J. F.
2007, ApJ, 664, 357

\refb Schultheis M., Ganesh S., Simon G. et al. 1999, A\&A, 349, L69

\refb Schultheis M., Glass I. S. 2001, MNRAS, 327, 1193

\refb Skrutskie M. F., Cutri R. M., Stiening R., Weinberg M. D. et al.
2006, AJ, 131, 1163

\refb Strai\v{z}ys V., \v{C}ernis K., Barta\v{s}i\={u}t\.e S. 1996,
Baltic Astronomy, 5, 125

\refb Strai\v{z}ys V., \v{C}ernis K., Barta\v{s}i\={u}t\.e S. 2003,
A\&A, 405, 585

\refb Strai\v{z}ys V., Corbally C. J., Laugalys V. 2008, Baltic
Astronomy, 17, 125

\refb Strai\v{z}ys V., Lazauskait\.e R. 2008, Baltic Astronomy, 17, 277
(this issue)

\refb Tiede G. P., Frogel J. A., Terndrup D. M. 1995, AJ, 110, 2788

\refb Wilking B. A., Gagn\'e M., Allen L. E. 2008, in {\it Handbook of
Star Forming Regions}, vol.\,2, ed.  B. Reipurth, ASP, p.\,351

\end{document}